\documentclass[a4paper,11pt]{article}
\pdfoutput=1 

\usepackage{jinstpub} 

\usepackage[T5,T1]{fontenc} 

\usepackage{siunitx}
\usepackage{todonotes}
\usepackage{gensymb} 
\usepackage{subcaption} 
\usepackage{booktabs} 
\usepackage{lineno} 
\usepackage{placeins} 

\renewcommand{\t}{\text} 

\title{A Convolutional Neural Network based Cascade Reconstruction for the IceCube Neutrino Observatory}

\author[16]{R. Abbasi,}
\author[56]{M. Ackermann,}
\author[17]{J. Adams,}
\author[11]{J. A. Aguilar,}
\author[21]{M. Ahlers,}
\author[47]{M. Ahrens,}
\author[27]{C. Alispach,}
\author[30]{A. A. Alves Jr.,}
\author[40]{N. M. Amin,}
\author[13]{R. An,}
% ...
\affiliation[11]{Universit{\'e} Libre de Bruxelles, Science Faculty CP230, B-1050 Brussels, Belgium}
% ...
\affiliation[13]{Department of Physics and Laboratory for Particle Physics and Cosmology, Harvard University, Cambridge, MA 02138, USA}
% ...
\affiliation[16]{Department of Physics, Loyola University Chicago, Chicago, IL 60660, USA}
\affiliation[17]{Dept. of Physics and Astronomy, University of Canterbury, Private Bag 4800, Christchurch, New Zealand}
% ...
\affiliation[21]{Niels Bohr Institute, University of Copenhagen, DK-2100 Copenhagen, Denmark}
% ...
\affiliation[27]{D{\'e}partement de physique nucl{\'e}aire et corpusculaire, Universit{\'e} de Gen{\`e}ve, CH-1211 Gen{\`e}ve, Switzerland}
% ...
\affiliation[30]{Karlsruhe Institute of Technology, Institute for Astroparticle Physics, D-76021 Karlsruhe, Germany }
% ...
\affiliation[40]{Bartol Research Institute and Dept. of Physics and Astronomy, University of Delaware, Newark, DE 19716, USA}
% ...
\affiliation[47]{Oskar Klein Centre and Dept. of Physics, Stockholm University, SE-10691 Stockholm, Sweden}
% ...
\affiliation[56]{DESY, D-15738 Zeuthen, Germany}
% ...

\emailAdd{analysis@icecube.wisc.edu}

\abstract{
Continued improvements on existing reconstruction methods
are vital to the success of high-energy 
physics experiments, such as the IceCube Neutrino Observatory. 
In IceCube, further challenges arise as the detector is situated at the geographic South Pole 
where computational resources are limited. 
However, to perform real-time analyses and to issue alerts to telescopes around the world, 
powerful and fast reconstruction methods are desired. 
Deep neural networks can be extremely powerful, and their usage is computationally inexpensive 
once the networks are trained. 
These characteristics make a deep learning-based approach an excellent candidate 
for the application in IceCube.
A reconstruction method based on convolutional architectures and hexagonally shaped kernels is presented.
The presented method is robust towards systematic uncertainties in the simulation and has 
been tested on experimental data. 
In comparison to standard reconstruction methods in IceCube, it can improve upon the reconstruction accuracy, while reducing the time necessary to run the reconstruction by two to three orders of magnitude.
}


\collaboration[c]{on behalf of the IceCube collaboration}

\begin{document}
\maketitle
\flushbottom


\section{Potential of Deep Learning-based Reconstruction Methods in IceCube}
\label{sec:intro}


The primary goal of the IceCube Neutrino Observatory is the detection of astrophysical neutrinos (neutrinos originating from extraterrestrial sources), which was achieved in 2013~\cite{HESE2yr}, as well as the identification and characterization of their sources. 
Although many efforts have been put forth to detect the sources of high-energy astrophysical neutrinos, to date, their origin remains a mystery.
However, first evidence for potential neutrino sources were found in observations of the blazar
TXS 0506+056~\cite{TXSMultimessenger, TXSArchival} and in time-integrated neutrino source searches~\cite{PS10yr}. 
Ongoing work is focused on improving detector calibration and reconstruction methods to further increase the sensitivity of the detector.
In addition, a real-time multi-messenger approach
is implemented to further increase the discovery potential through the combination of measurements from different telescopes~\cite{RealTimeAlertSystem,IceCubeMagicVeritasFollowUp}.

High-energy neutrino candidates are selected and reconstructed on-site at the South Pole.
Given the hardware limitations, events must be processed within a certain amount of time to prevent pileup. 
Therefore, only simple and fast event reconstructions can be run on-site.
Even though more sophisticated and powerful reconstruction methods exist~\cite{DirectFit, EarlyMuonCascadeReco}, these can take minutes to hours to reconstruct a single event, rendering them intractable for use at the South Pole system.
The high data rate as well as the strict hardware and bandwidth limitations are key challenges to the success of the real-time follow-up framework.
But even for off-site reconstructions, the most advanced methods in IceCube can only be applied to a subset of data due to their computational complexity. 
Offline analyses can therefore also benefit from improved and more efficient reconstruction methods.

These limitations call for a robust reconstruction method which can 
handle raw data and has a short and preferably constant runtime. 
Any new reconstruction method would ideally also reduce the uncertainties on 
reconstructed parameters in comparison to the standard reconstruction method.
Recent advances in image recognition~\cite{ImageNetBreakthrough} have shown the capabilities of deep convolutional networks to surpass state-of-the-art methods.
Once the networks are trained, their usage is computationally inexpensive.
The network performs a fixed amount of mathematical operations on the input data, resulting in a stable runtime that is essentially independent of the input.
These characteristics make deep learning-based methods an excellent candidate for powerful and fast on-site reconstructions.

In this paper, a reconstruction method based on convolutional neural networks (CNNs) is presented which utilizes
hexagonally shaped convolution kernels.
The neural network is set up as a multi-regression task.
It is capable of achieving competitive reconstruction accuracy on 
neutrino properties, such as their incoming direction and energy, and 
estimating per-event uncertainties on these quantities. Simultaneously, 
the neural network achieves runtimes that are reduced by two to three 
orders of magnitude in comparison to standard reconstruction methods in 
IceCube.
The reconstruction method is validated by investigating the effects of systematic uncertainties in the simulation and by demonstrating agreement between simulated and experimental data.
It has been applied in Ref.~\cite{CascadePaperMike7yrs} with minor modifications to the network architecture and training procedure.
Related work and applications in IceCube are discussed in Refs.~\cite{ICRC17DeepLearning, NERSC_GNN, IceCubeMunichClassification}.
Apart from IceCube, many other high-energy physics experiments have adopted deep learning-based reconstruction methods.
Applications utilizing CNNs include, but are not limited to,
work detailed in 
Refs.~\cite{NOvACNN, MicroBooNE01, DeepLearningLHC, DeepLearningNeutrinoExperiments, HESSCNNs, AUGERCNN, CNNKM3Net}.

%

The paper is structured as follows: 
in Section~\ref{sec:cascades}, cascades are defined and the IceCube detector is introduced.
Section~\ref{sec:data} and~\ref{sec:architecture} illustrate the data input format of the neural network and its architecture.
The training procedure is explained in Section~\ref{sec:training} and Section~\ref{sec:performance}
highlights the performance of the presented method.
Effects of systematic uncertainties are investigated in Section~\ref{sec:systematics}.
Before concluding in Section~\ref{sec:conclusion}, limitations of the chosen network architecture
are discussed in Section~\ref{sec:limitations}.

\section{Cascade Reconstruction in IceCube}
\label{sec:cascades}

The reconstruction method presented in this paper is a versatile tool that is
applicable to a wide range of classification and regression tasks.
In the scope of this paper, the method is demonstrated for the
reconstruction of cascade events.
In the following, the IceCube detector is introduced, cascade events
are defined, and the current standard cascade reconstruction method is discussed.

\subsection{The IceCube Detector}
\label{sec:cascades_icecube}

IceCube is a neutrino detector located at the South Pole instrumenting a cubic kilometer of glacial ice.
The detector consists of 5160 digital optical modules (DOMs) with a downward-facing {10~inch-diameter} photomultipler tube (PMT)~\cite{IceCubePMT} installed on 86 vertical strings at depths between $\SI{1450}{\m}$ and $\SI{2450}{\m}$.
The origin of the IceCube coordinate system is placed in the detector center at a depth of $\SI{1948}{\m}$.
The $z$-axis is chosen to point upwards towards the ice surface.
PMT signals are digitized and buffered on the DOM mainboard with a timing resolution of about 2ns.
Upon readout request, these digitized waveforms are sent to computers on the surface of the detector array.
%
%
\begin{figure}
  \begin{subfigure}[b]{0.6\textwidth}
    \includegraphics[width=\textwidth, keepaspectratio]{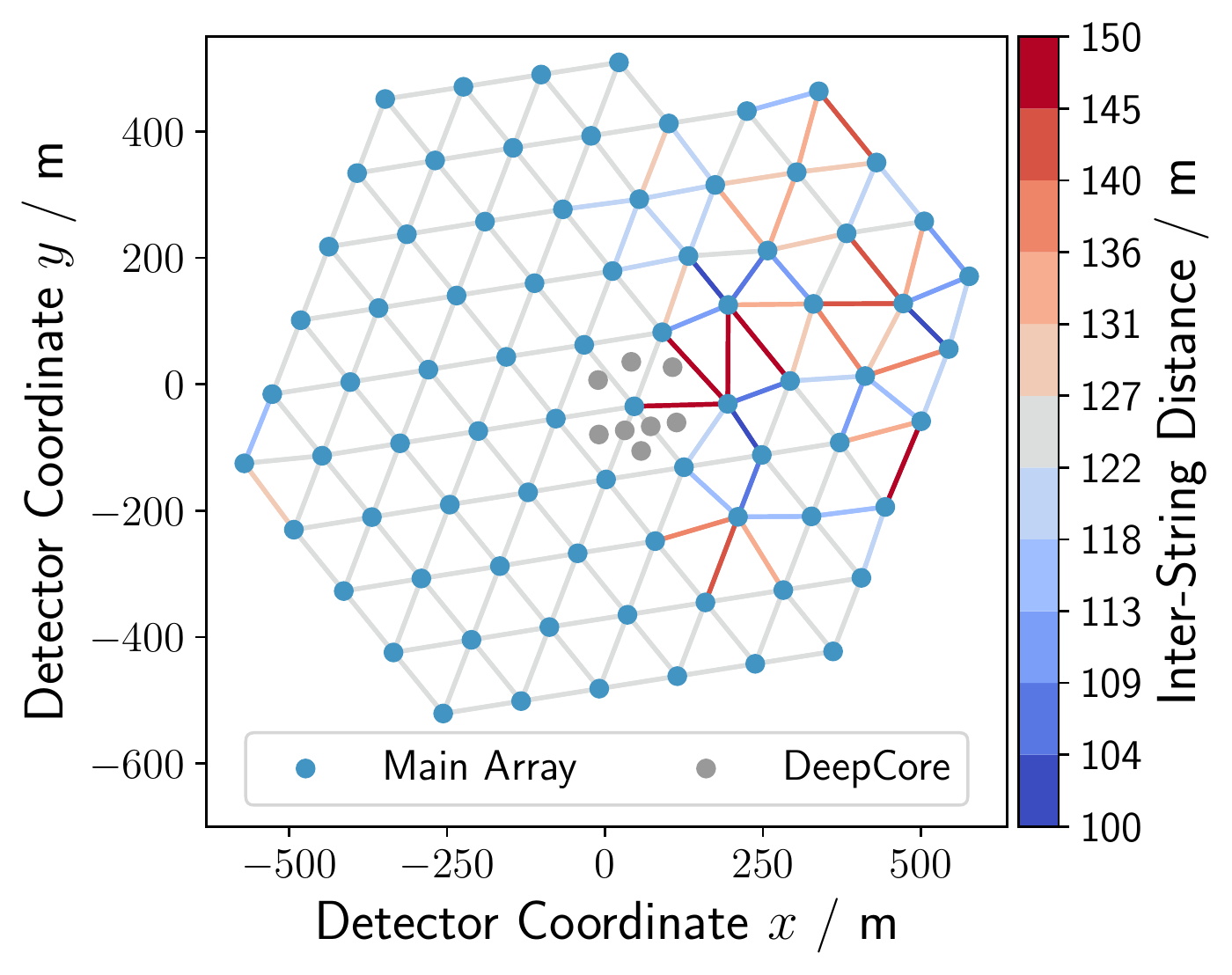}
    \caption{Top view of IceCube}
    \label{fig:detector_grid_x_y}
  \end{subfigure}
  \hfill
  \begin{subfigure}[b]{0.34\textwidth}
    \includegraphics[width=\textwidth, keepaspectratio]{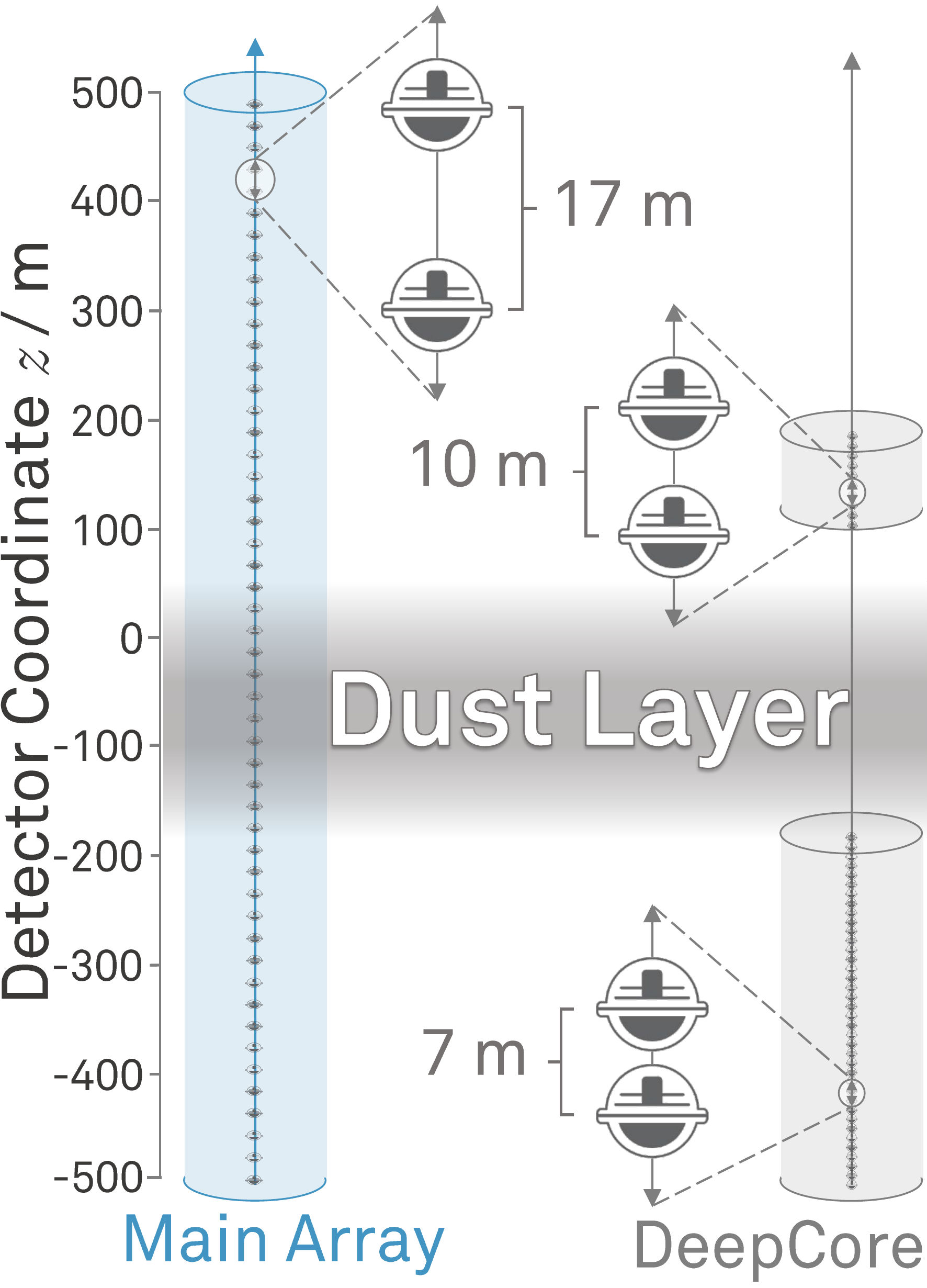}
    \caption{Side view of Strings}
    \label{fig:detector_grid_z}
  \end{subfigure}
  \caption{
  A top view of the IceCube detector is shown on the left. The in blue
depicted 78 strings are on an approximately triangular grid, while the DeepCore
strings, shown in gray, are installed in a denser configuration. 
The color scale indicates the inter-string distance for the main IceCube array, 
which can substantially deviate from the usual spacing of $\SI{125}{m}$.
On the right, the DOM layout along the $z$-axis of the strings is illustrated.
In contrast to the strings of the main array, the DOMs on the DeepCore strings (gray) are divided into two groups,
one above the dust layer and one below.
  }
  \label{fig:detector_grid}
\end{figure}
The 86 strings of the detector may be grouped into three detector parts.
While the main IceCube array, consisting of the first 78 strings, is arranged on an approximately triangular grid, the remaining 8 strings do not follow this symmetry.
They are installed in a denser configuration called DeepCore~\cite{DeepCore} with variable distances to neighboring strings.
Each of the 86 strings holds 60 DOMs.
These are evenly distributed along the $z$-axis for strings of the main array.
DOMs located on DeepCore strings are further grouped into an array above the dust layer (10 DOMs on each string) and an array below the dust layer (50 DOMs per string) as illustrated in Fig.~\ref{fig:detector_grid_z}.
The dust layer is a layer in the glacial ice with increased dust impurities that was produced about 60000 to 70000 years ago~\cite{IceCubeDustLayer, IceProperties}. 
It ranges from depths of $\SI{1950}{m}$ to $\SI{2100}{m}$ and results in increased scattering and absorption coefficients.
The strings of the main IceCube array have an inter-string spacing of about $\SI{125}{\meter}$. 
However, there are deviations as shown in Fig.~\ref{fig:detector_grid_x_y}.

At trigger level, IceCube data consist of recorded waveforms
where the amplitude of the waveform corresponds to the charge recorded by the DOM.
An IceCube 'event' is a set of these waveforms for many DOMs.
Each DOM can measure multiple waveforms with variable starting times in a single event.
While a typical event has a read out window of approximately $\SI{15000}{\nano\second}$, this can vary depending on the trigger~\cite[pp.56-59]{DetectorPaper}.
In a subsequent step, a series of  pulses, represented as a series of times and
amplitudes of light observed in a DOM, are extracted from the recorded waveforms.
The number of extracted pulses can vary by orders of magnitude and it is different for each DOM.
A detailed description of the IceCube detector and its data acquisition system can be found in Ref.~\cite{DetectorPaper}.
\subsection{Cascade Events}
\label{sec:cascades_cascades}

\begin{figure}
  \begin{subfigure}[b]{0.49\textwidth}
    \includegraphics[width=\textwidth, keepaspectratio]{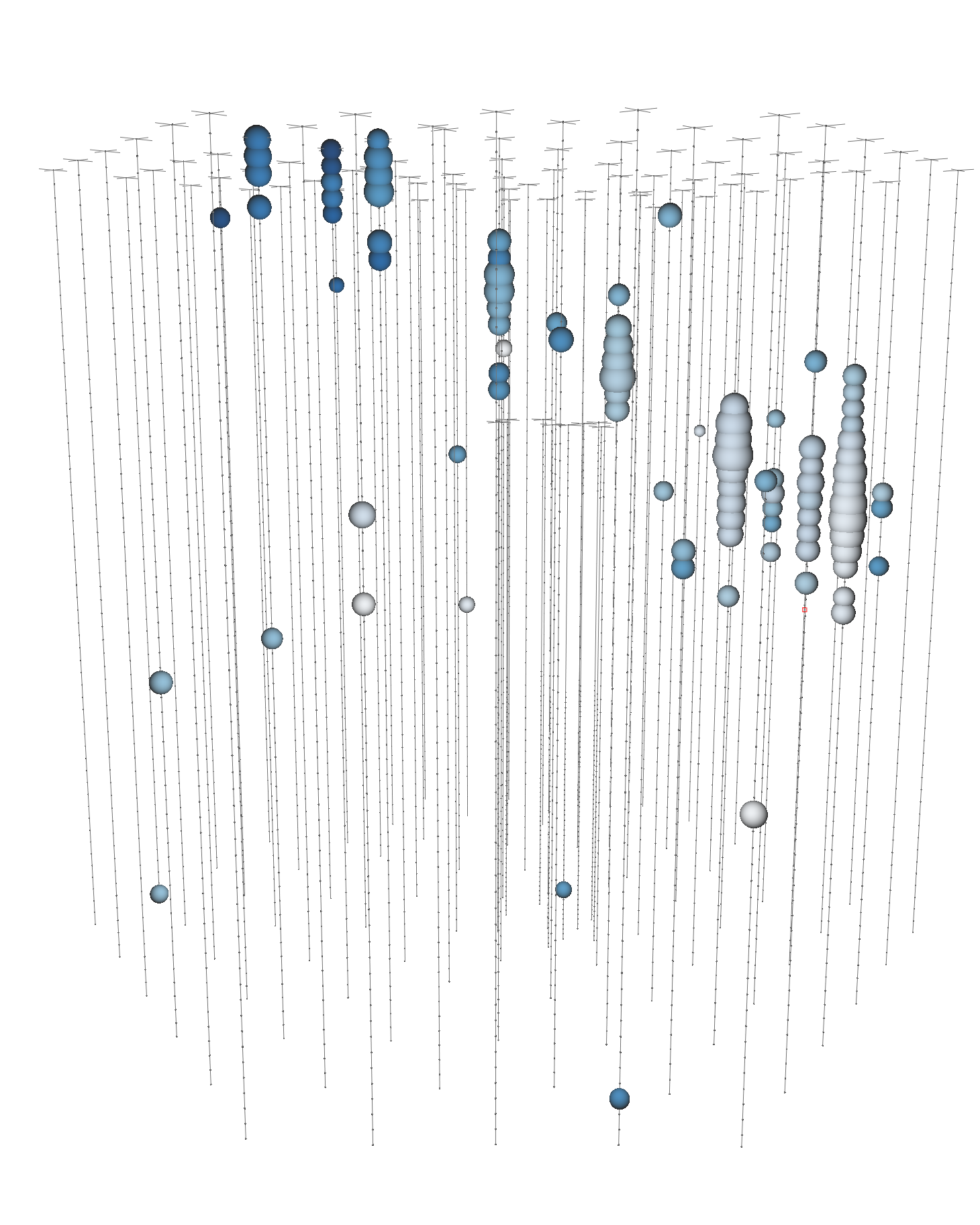}
    \caption{Up-going muon from CC $\nu_\mu$ interaction}
    \label{fig:eventview_numu}
  \end{subfigure}
  \begin{subfigure}[b]{0.49\textwidth}
    \includegraphics[width=\textwidth, keepaspectratio]{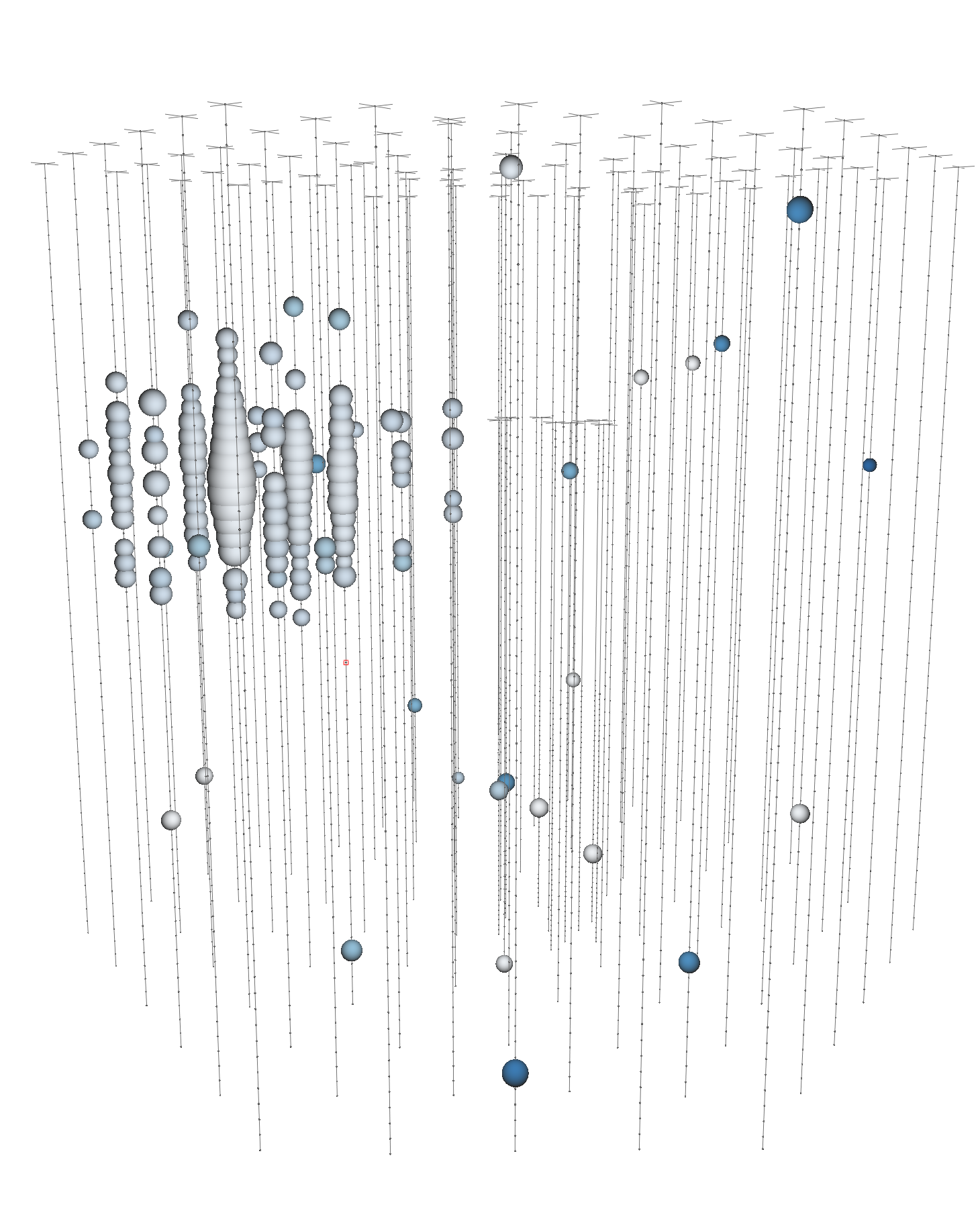}
    \caption{CC $\nu_e$ interaction inside detector volume}
    \label{fig:eventview_nue}
  \end{subfigure}
  \caption{
  Example event views of simulated data are shown for an up-going muon entering the detector (left)
  and for the resulting particle cascade of a charged-current (CC) $\nu_e$ interaction inside the detector volume (right). 
  Each DOM is represented by a sphere. The size of the sphere corresponds to the amount of collected light and the color indicates the arrival time of the photons (darker colors correspond to later times).
  }
  \label{fig:eventview}
\end{figure}

In order to detect neutrinos, IceCube measures Cherenkov photons produced by charged secondary particles resulting from neutrino interactions.
The two primary detection channels consist of so called track-like events (tracks), muons induced by charged-current~(CC) $\nu_\mu$ interactions,
as well as cascade-like events (cascades), which result from CC $\nu_e$ and $\nu_\tau$ interactions in addition to neutral current interactions of all neutrino types~\cite[p. 6] {RealTimeAlertSystem}.
Muons deposit energy along their trajectory through the detector resulting in tracks as shown
in Fig.~\ref{fig:eventview_numu}.
Cascade-like events, on the other hand, produce a shower of secondary particles at the neutrino
interaction vertex, which is generally not resolvable by IceCube, given its inter-string spacing of
about~$\SI{125}{\meter}$ (see Section~\ref{sec:cascades_icecube}).
As a result, IceCube detects a spherical, almost point-like energy deposition as shown in Fig.~\ref{fig:eventview_nue}.


In contrast to track-like events, the missing lever arm and the spherical energy deposition make
the angular reconstruction of cascades a challenging task in IceCube.
Furthermore, cascade reconstructions are more susceptible to local ice properties than 
track reconstructions due to the local energy deposition of the cascade.
For tracks, the energy depositions are distributed over large parts of the detector, which
helps to average out local fluctuations.
It is therefore crucial to understand the effect that systematic uncertainties in the description of
the local ice properties may have on the reconstruction (this is investigated in Section~\ref{sec:systematics}).
Despite these challenges, cascade events are an important detection channel that expand 
IceCube's ability to probe the southern neutrino sky, which is otherwise dominated by down-going atmospheric muons.

The main cascade reconstruction quantities of interest are the deposited energy and the
direction of the primary neutrino that induced the particle shower.
The focus of this paper is on the angular reconstruction of cascade-like events in the TeV to PeV energy range.

\subsection{Standard Cascade Reconstruction Method}
\label{sec:cascades_reco}

To combat the challenges involved in the reconstruction of cascades,
IceCube has implemented a variety of reconstruction algorithms.
The current standard cascade reconstruction method is a 
maximum likelihood estimation that fits a point-like cascade template to the measured light deposition pattern in the detector~\cite{EnergyReconstruction}.
The cascade template describes the expected light yield at each DOM for a given cascade hypothesis
consisting of the interaction vertex and time, the deposited energy, and the direction of the
primary neutrino.
A binned Poisson likelihood is employed to find the cascade hypothesis that best describes
the measured light deposition pattern in the detector.
The cascade template is obtained from MC simulation and tabulated for various cascade-DOM-configurations.
In order to reduce the dimensionality of the lookup tables, 
rotational and translational invariance in the $x$-$y$-plane are assumed. 
Second order corrections can then be applied to account for the inhomogeneities in the 
detector medium and photon propagation. 
Due to these simplifications, 
the standard reconstruction method maximizes an approximation to the true underlying likelihood.
More advanced reconstruction methods exist within IceCube~\cite{DirectFit, EarlyMuonCascadeReco}, however, due to their computational complexity they can only be applied on single events.
Additional information on the standard reconstruction method is provided in Ref.~\cite{EnergyReconstruction}.

\section{Neural Network Input Data Format}
\label{sec:data}


Convolutional neural networks (CNNs) were first developed for the domain of image recognition~\cite{ConvolutionalNetworks}.
In contrast to fully-connected (dense) networks, CNNs can exploit translational invariance in the input data.
Translational invariance in the context of image recognition means that the class of an image does not
change if the position of the classified object is shifted within that image.
The capability of CNNs to exploit translational invariance is one of several contributions that led to the breakthrough in image recognition~\cite{ImageNetBreakthrough}.
In the case of IceCube, the application of CNNs may help in exploiting the fact that the underlying physics of the neutrino interaction are invariant under translation in space and time.
Inputting image data into a convolutional neural network is straight forward, as it only has two dimensions with an optional color channel.
The pixels of an image are arranged on an orthogonal grid and for typical image datasets the pictures are all of the same size or can easily be cropped to the same size.
In contrast, IceCube data is given at four dimensional points in space and time, arranged on an imperfect triangular grid and highly variable in size.
It has to be unified in a way that it can be sequenced into a convolutional neural network while maintaining
spatial and temporal relations so that symmetries, such as translational invariance in space and time,
can be exploited.
Key challenges that need to be addressed include the high dimensionality and variability of the data as well as IceCube's hexagonal geometry and sub-arrays.

\subsection{IceCube's Hexagonal Geometry and Sub-arrays}
\label{sec:data_input_tensors}

For purposes of use in the CNN, the IceCube detector is effectively separated
into three detector parts, which do not
share the same geometry or symmetries. The main IceCube array is arranged on an
approximately triangular grid. However, pixels in an image are typically arranged
on an orthogonal grid. Data storage, matrix and tensor operations are also performed
on orthogonal grids.
In order to employ existing and efficient convolution algorithms,
IceCube data must first be transformed to an
orthogonal grid.
A common approach 
to transform hexagonal data 
is to interpolate or rebin the data points 
onto an orthogonal grid as has been done 
elsewhere~\cite{HEXSampling04_VERITAS, HEXSampling03_HESS, HEXSampling01_CTA}.
However, this process introduces an unnecessary loss of information as the original data points on the hexagonal grid are replaced with an approximation thereof. 
In addition, while the interpolation is reasonable for the integrated charge, 
it is not clear how this interpolation would have to be performed on the pulse arrival times 
which heavily depend on the 
event characteristics and cannot generally be inferred through the measured distribution of pulses at DOMs in the vicinity.

\begin{figure}
  \centering
  \includegraphics[width=0.8\textwidth, keepaspectratio]{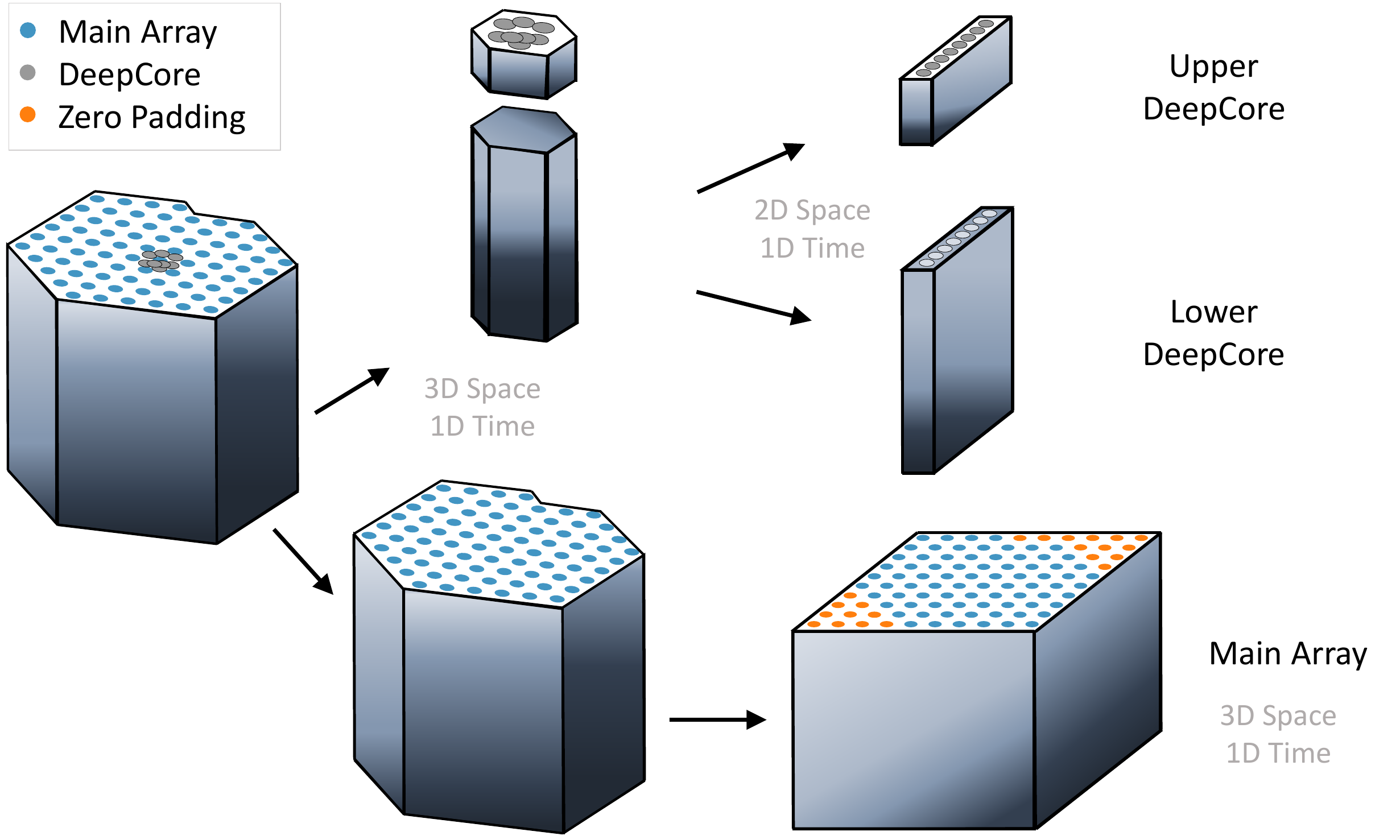}
  \caption{The main IceCube array and DeepCore strings are handled separately due to their differing geometry. Hexagonally shaped data of the main array can be transformed from an axial coordinate system into an orthogonal grid by padding with zeros (orange dots) and aligning the rows, which results in a $10\times 10$ grid in the $x$-$y$-plane. Every DOM defines a bin in the spatial coordinates (DOM-binning).
    } \label{fig:IceCubeDataInput}
\end{figure}

Instead, a lossless transformation is used~\cite{ICRC17DeepLearning}, which is also employed in Section~\ref{sec:architecutre_residuals} to obtain hexagonal convolution kernels.
In this approach, each DOM itself is used as a bin and input node for the CNN.
For this DOM-binning, the triangular grid of the DOMs needs to be transformed to an orthogonal grid.
This can be achieved by the transformation described in Fig.~\ref{fig:IceCubeDataInput}.
In order to exploit the symmetries maintained within each sub-array of the detector,
the two DeepCore parts of the detector are handled separately from the main array, reducing the $x$ and $y$-coordinates to a single dimension. 
Note that this separate processing results in a loss of ability to directly correlate neighboring strings
from the main array with those of DeepCore.
Furthermore, the transformation as outlined above does not explicitly correct
for the dust layer or for the irregularities in the detector grid, detailed in Section~\ref{sec:cascades_icecube}. 
%
%
%
As a result of the transformation, three input tensors of the shape $(10\times 10\times 60\times n)$, $(8\times 10\times n)$, and $(8\times 50\times n)$ are obtained for the main array and upper and lower DeepCore, respectively, where $n$ denotes the number of input variables per DOM.
The second to last dimension in each tensor indicates the number of DOMs (60, 10, and 50) along the z-axis.
Possible input variables for each DOM are discussed in the following section.

\subsection{Data Dimensionality and Variability}
\label{sec:data_dom_inputs}

The transformation described above allows for a convolution over all spatial dimensions for the main array and a convolution over the z-dimension for DeepCore.
Hence, the symmetry in spatial coordinates can be exploited to the extent possible given IceCube's geometry.
Ideally, translational invariance in time should also be exploited.
The starting point of most reconstruction methods in IceCube is the extracted pulses as described in Section~\ref{sec:cascades_icecube}.
These pulses are an efficient data representation.
However, the number of pulses at a given DOM is highly variable and therefore the pulses are an unsuitable input to a CNN.
A standard CNN requires a uniform and constant input size.
One option is to bin the measured pulse charges in time.
With a four dimensional convolution over the main array input, 
translational invariance in space and time 
can be exploited.
A full four dimensional convolution would require on the order of thousands of time bins per DOM for the desired timing resolution.
This would result in a significant increase in computational complexity.
A simultaneous processing in space and time via a 4D convolution is therefore not feasible.
The dimensionality of the problem must be reduced or separated out in individual tasks.
Reducing the dimensionality of the problem results in a loss of spatial and temporal relations that have to be compensated for in an alternative way.

%
%
\begin{figure}
  \centering
  \includegraphics[width=\linewidth, keepaspectratio]{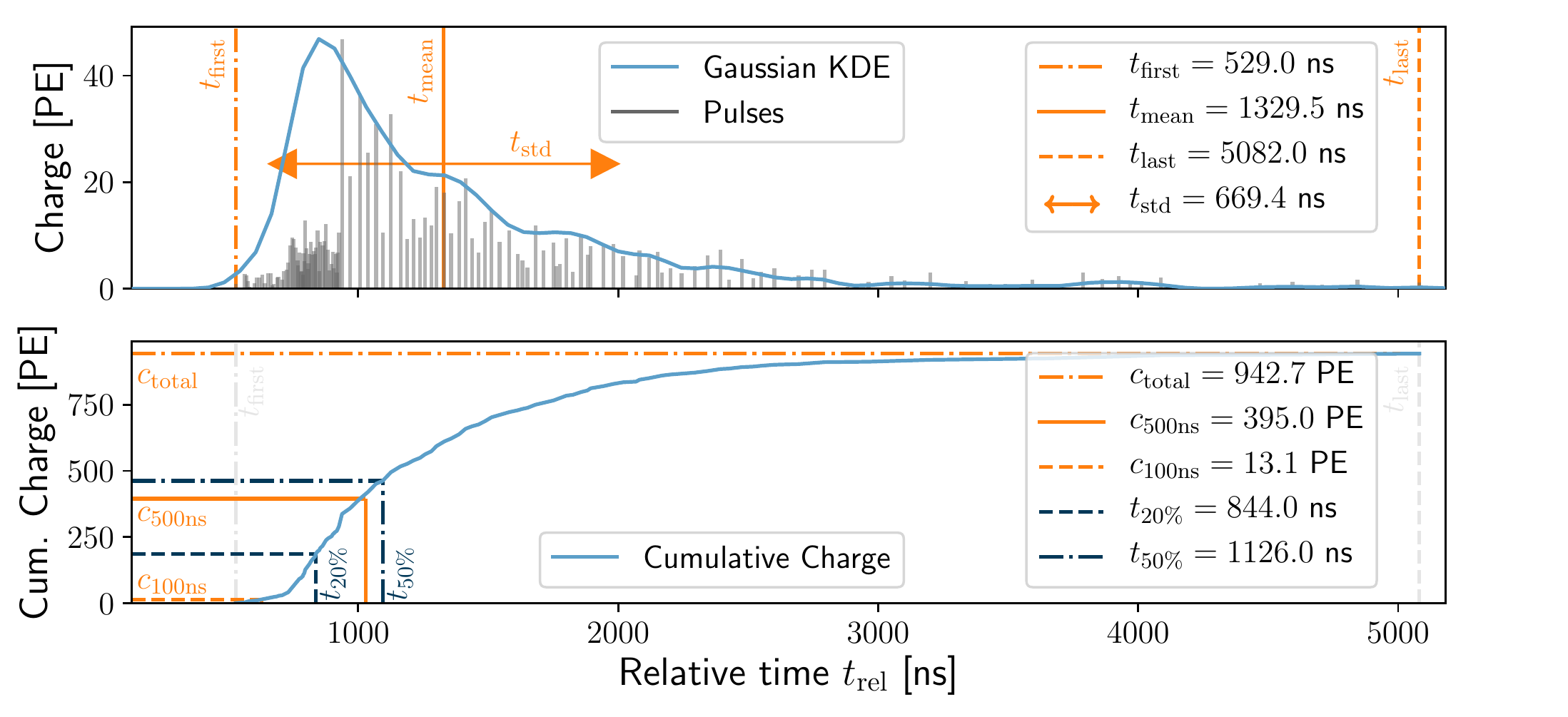}
  \caption{
  An example pulse series and corresponding input data for a single DOM is shown.
  The measured pulse series cannot directly be utilized by a CNN due to its varying length.
  The pulses are therefore reduced to nine input parameters ($c_\mathrm{total}$, $c_\mathrm{500ns}$, $c_\mathrm{100ns}$, $t_\mathrm{first}$, $t_\mathrm{last}$, $t_\mathrm{20\%}$, $t_\mathrm{50\%}$, $t_\mathrm{mean}$, $t_\mathrm{std}$) which aim to summarize the pulse distribution.
  } 
\label{fig:wf_input_data}
\end{figure}
%
%

There are many ways to achieve this.
For this paper, the variable size of the time dimension is reduced to nine selected summary statistics, which describe the pulse series at each DOM.
This has the added benefit that parameters can be chosen that have good agreement between measured and simulated data, making the network more robust towards possible mis-modeling in the Monte Carlo (MC) simulation.
Parameters including time variables are calculated relative to a global offset which is defined as the start of a $\SI{6000}{ns}$ long time window that maximizes the contained charge for each event.
By utilizing relative timing information rather than absolute timing, 
translational invariance in time can be exploited to a certain degree.
The nine input parameters are chosen based on their expected relevance for the reconstruction task and consist of:
the total DOM charge,
the charge within 100~ns and 500~ns of the first pulse,
the relative time of first pulse,
the relative time at which 20\% and 50\% of the charge is collected,
the relative time of the last pulse,
and the charge weighted mean and standard deviation of the relative pulse arrival times.
The input parameters are illustrated in Fig.~\ref{fig:wf_input_data}.
Reconstruction methods in IceCube typically exclude pulses from saturated or overly bright DOMs to avoid potential mis-modeling in the MC simulation.
In this case, the input features corresponding to the excluded DOMs are set to zero.
With these features, three input tensors are obtained for the CNN of the shape $(10\times 10\times 60\times 9)$, $(8\times 10\times 9)$, and $(8\times 50\times 9)$ for the main array and upper and lower DeepCore, respectively.

In a future iteration, the calculation and selection of input parameters (feature generation) for a given DOM could be automated by applying a 1D CNN on the measured waveforms or time-binned pulses. 
Alternatively, a recurrent neural network can be set up to work directly on the extracted pulses.
The automated and fully differentiable feature generation would allow for an end-to-end learning task starting from the measured waveforms or pulse series.

\section{Neural Network Architecture and Design Choices}
\label{sec:architecture}

Once the data input format is defined, the network architecture can be set up.
The reconstruction quantities of interest (training labels) are defined as the 
deposited energy in the detector and the direction of the 
incoming neutrino.
The zenith~$\Theta$ and azimuth~$\Phi$ angle of the neutrino direction is further decomposed into the
direction vector components.
Due to a widely used convention in IceCube,
the particle direction vector~$\vec{d}$ is chosen to point in the direction of travel while
$\Theta$ and~$\Phi$ define the origin of the particle.
This results in a total of 6 labels which consist of: 
energy, azimuth, zenith, dir.-$x$, dir.-$y$, dir.-$z$.
Settings and design choices of the neural network are discussed in this section.
The neural network is implemented within the python interface of \textsf{TensorFlow}~\cite{tensorflow2015-whitepaper} and the code is available on GitHub\footnote{\url{https://github.com/icecube/dnn_reco}}.

\begin{figure}
  \centering
  \includegraphics[width=\linewidth, keepaspectratio]{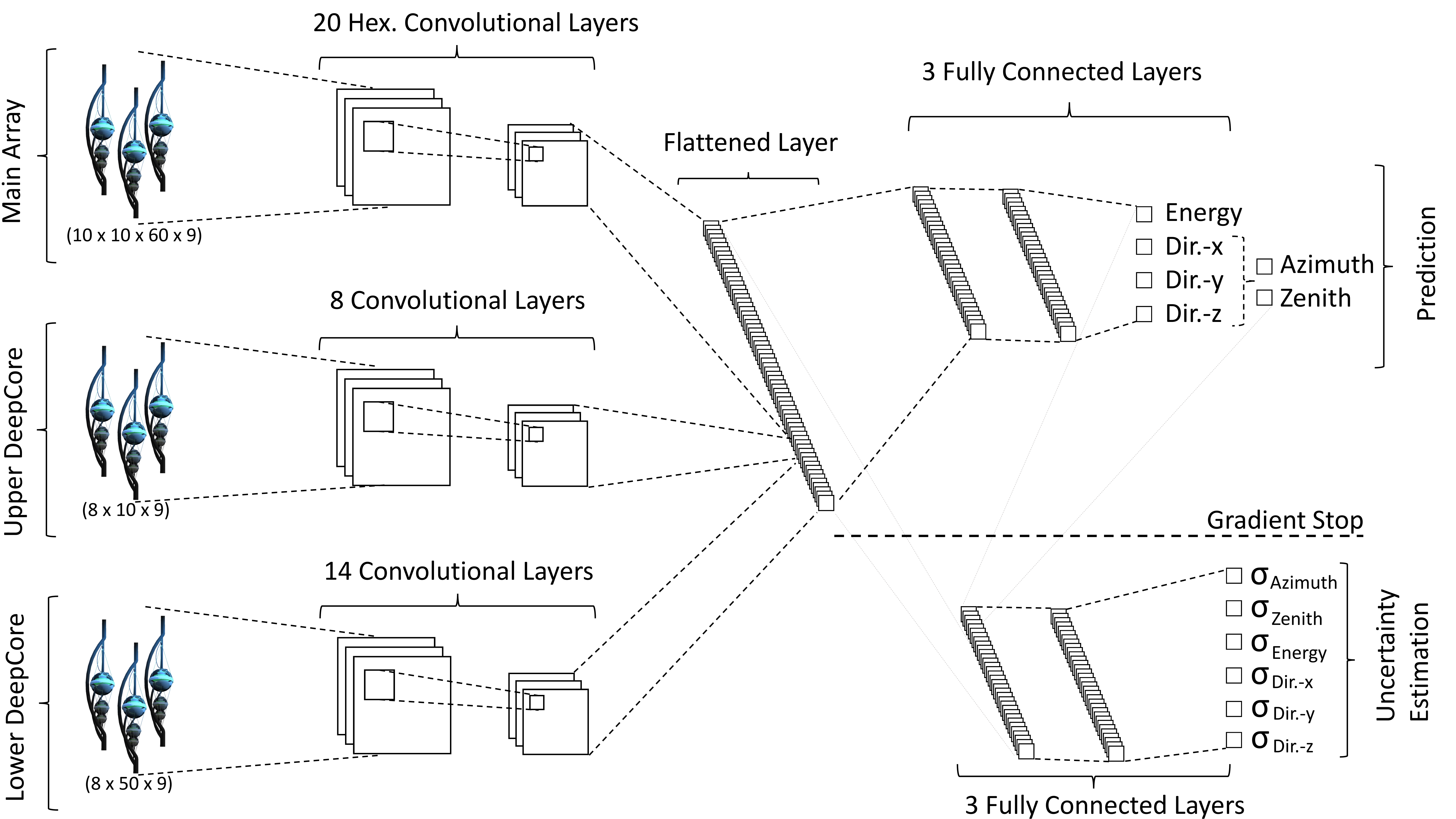}
  \caption{A sketch of the neural network architecture is shown. Data from the three sub-arrays are sequenced into convolutional layers. The result is flattened, combined, and passed on to two fully-connected sub-networks which perform the reconstruction and uncertainty estimation.
  The uncertainty-estimating sub-network also obtains the prediction output as an additional input.
  }
  \label{fig:architecture}
\end{figure}

\subsection{Hexagonal Convolution Kernels}
\label{sec:architecture_hex}

\begin{figure}
  \centering
  \includegraphics[width=\linewidth, keepaspectratio]{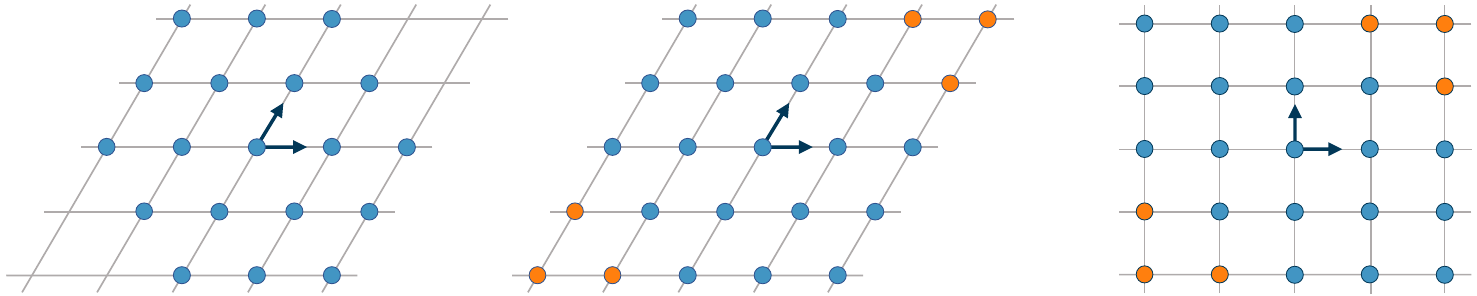}
  \caption{Hexagonally shaped convolutional kernels in an axial coordinate system on the left (blue dots) can be transformed into an orthogonal grid on the right by padding with zeros (orange dots) and aligning the rows~\cite{ICRC17DeepLearning}.}
  \label{fig:hexKernel}
\end{figure}

\begin{figure}
  \begin{minipage}[t]{0.70\textwidth}
    \vspace{0pt}
    \includegraphics[width=\linewidth, keepaspectratio]{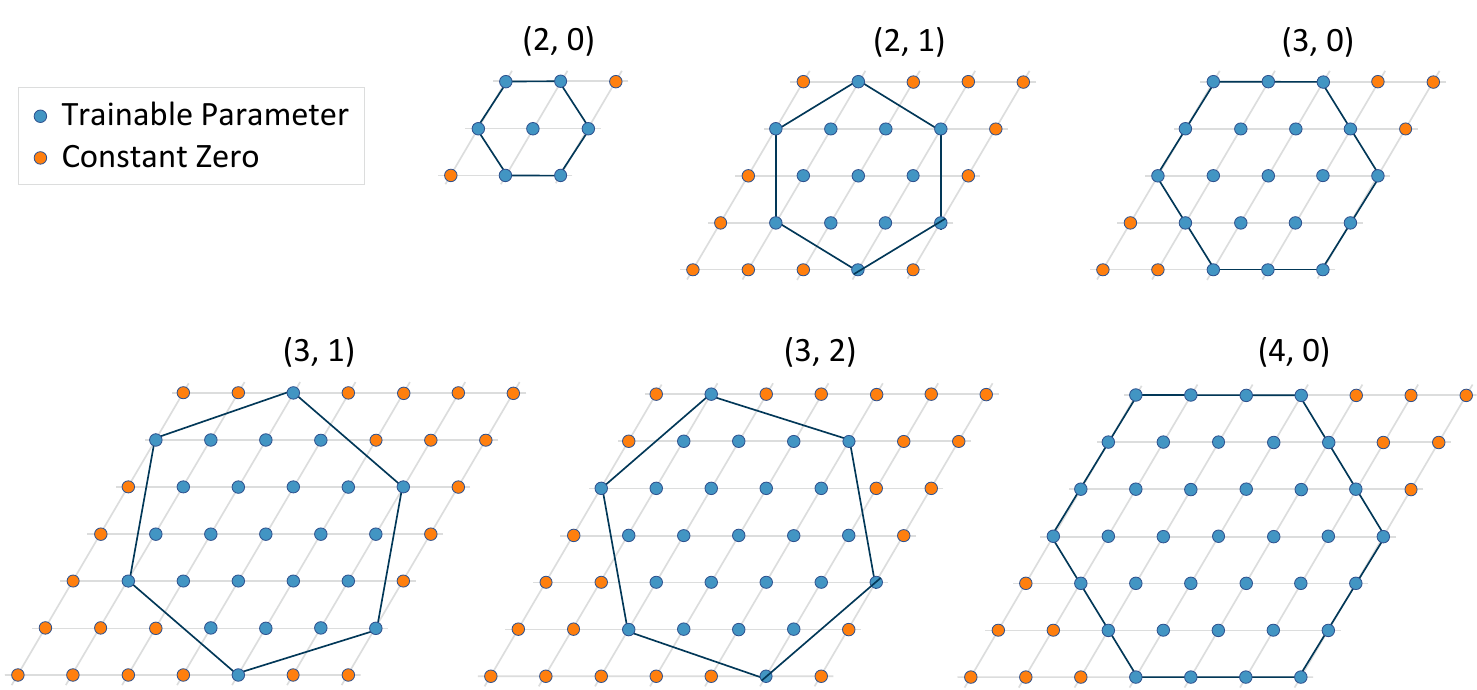}
  \end{minipage}\hfill
  \begin{minipage}[t]{0.28\textwidth}
    \vspace{5pt}
    \caption{An hexagonally shaped kernel can be defined by a tuple of size~$s$ and orientation~$o$: $(s,o)$. For an axis aligned hexagon (orientation~$o=0$), the size parameter~$s$ defines the number of points along an edge of the hexagon.}
  \label{fig:HexKernelSizeOverview}
  \end{minipage}
\end{figure}

Convolution operations in common deep learning frameworks, such as \textsf{TensorFlow}, are performed on orthogonal grids.
Transforming the (in $x$-~and $y$-dimension) hexagonally shaped IceCube data to an orthogonal grid as illustrated in Fig.~\ref{fig:IceCubeDataInput} results in convolution kernels shaped as parallelograms in the detector.
To instead obtain convolutional kernels which are shaped according to the IceCube geometry,
the transformation method described in our previous work~\cite{ICRC17DeepLearning} is applied.
The convolution kernels within \textsf{TensorFlow} are adjusted by setting the corner elements to zero as illustrated by the orange points on the right of Fig.~\ref{fig:hexKernel}.
%
As a result, hexagonally shaped convolution kernels are obtained for the $x$-$y$-plane in the detector.
The obtained kernels can be defined by a tuple of size and orientation as illustrated in Fig.~\ref{fig:HexKernelSizeOverview}.
Due to their symmetry, these kernels can in principle be rotated by multiples of $\SI{60}{\degree}$ to exploit rotational equivariance about the $z$-axis within group convolutions as explained in Refs.~\cite{GroupConvolutions, GroupConvolutionsHex}.
%

\subsection{Data Normalization}
\label{sec:architecture_data}

In order for neural networks to be able to learn non-linear relations, non-linear functions (activation functions) are applied to the output of a layer.
Once the architecture of the neural network is defined, the parameters of the network can be adjusted to minimize a specified target function, 
typically referred to as loss function or loss.
The loss function 
quantifies the deviation from the target values. 
It is used to determine which set of parameter values
best solves the reconstruction task.
During the training process, the loss is minimized iteratively via a gradient-based minimizer~\cite{ADAMOptimizer}.
Details on the chosen loss function are provided in Section~\ref{sec:training_loss}.

Deep neural networks can in theory process data of any range and scaling.
However, the activation function's response is typically centered around zero and often converges
for very large or small values leading to vanishing gradients.
Moreover, the loss landscape's dependence on the parameters of the neural network may be highly asymmetric if input features are on different size scales, 
i.e. if the order of magnitude between the features differ,
which can cause problems during minimization.
In addition, if input data is not normalized, small imbalances in the early layers of the network can cascade throughout neural networks causing gradients to become extremely large during backpropagation.
This effect, often referred to as exploding gradients, is even stronger for deep neural networks~\cite{EfficientBackprop, SelfNormalizingNN}.

It is therefore useful to normalize the input data~$X$ and training labels~$Y$.
Note that~$X$ and~$Y$ are tensors. 
Hence, the following transformations, outlined in Eqns.~\eqref{eqn:log} and~\eqref{eqn:normalization}, are applied pointwise.
IceCube measures neutrino interactions across many orders of magnitude in energy.
As a result, the energy label and input features such as the collected charge span over many orders of magnitude.
Consequently, these input features and labels are first transformed via
\begin{equation}
X' = \ln\left( 1.0 + X \right) \quad \text{and} \quad
Y' = \ln\left( 1.0 + Y \right).
\label{eqn:log}
\end{equation}
The transformation defined in Eq.~\eqref{eqn:log} only affects the input features total charge, charge collected within the first $\SI{100}{ns}$ and $\SI{500}{ns}$, as well as the energy label.
The identity operation $X'=X$ and $Y'=Y$ is applied to all other features and labels, respectively.
Afterwards, the input data~$X'$ and labels~$Y'$ are normalized to zero mean and unit variance by the following transformation:
\begin{equation}
X'' = \frac{X' - \overline{X'}}{\sigma_{X'} + \epsilon} \quad \text{and} \quad
Y'' = \frac{Y' - \overline{Y'}}{\sigma_{Y'} + \epsilon} \,\,\, , 
\label{eqn:normalization}
\end{equation}
where $\overline{X'}$ is the mean and $\sigma_{X'}$ is the standard deviation of $X'$ computed over all DOMs and over all events of the training dataset.
Entries in $\overline{X'}$ and $\sigma_{X'}$, which are a result of the zero-padding described in Fig.~\ref{fig:IceCubeDataInput}, are set to zero.
Therefore, padded zeros in the input data~$X$ will remain zero in the transformed input data~$X''$.
The tensors $\overline{Y'}$ and $\sigma_{Y'}$ are defined analogously for the labels.
A small constant $\epsilon = 10^{-4}$ is added to prevent division by zero.
The values $\overline{X'}$, $\sigma_{X'}$ and $\overline{Y'}, \sigma_{Y'}$ in Eq.~\eqref{eqn:normalization} are calculated in advance on the training dataset prior to the training of the neural network.


In addition to the transformations described by Eqns.~\eqref{eqn:log} and~\eqref{eqn:normalization}, steps are undertaken to maintain a standardized input to every layer throughout the network.
This can be accomplished by the use of residual additions 
and unit variance maintaining layers as described in Section~\ref{sec:architecutre_residuals} and~\ref{sec:architecture_variance}, respectively.
Other common approaches not regarded here are the application of self-normalizing networks~\cite{SelfNormalizingNN}
and batch normalization~\cite{BatchNormalization}.
Batch normalization is not used due to its regularization properties, which are too strong for the application at hand (see Section~\ref{sec:architecture_regularization}).


\subsection{Residual Additions}
\label{sec:architecutre_residuals}

Residual nets~\cite{ResidualNets} can greatly facilitate the training of deep neural networks by avoiding vanishing or exploding gradients through the introduction of skip connections.
Intuitively, the introduction of skip connections reduces the dependence among the layers and thus the cascading effect in deep neural networks.
The concept is adopted here with slight modifications applied. 
In residual layers as defined in this work, the output of a given layer~$X_\t{output}$ is modified to~$X'_\t{output}$ according to
\begin{equation}
  X'_\t{output} = X_\t{input} + s\cdot X_\t{output}
  \label{eqn:ResAddition}
\end{equation}
where $s$ is a scale factor defining the contribution of the residual of the given layer.
This allows the neural network to essentially skip the weights of a given layer by setting $s$ to zero.
The scale factor~$s$ is initialized to small random values.
As a result, the given layer will initially perform an approximate identity operation.
During training, the scale factor~$s$ is optimized and the contribution of the residual may increase. 
Due to downsizing or pooling operations in convolutional layers, the dimensions of $X_\t{input}$ and $X_\t{output}$ can differ.
In this case, only the channels which align are treated as residuals according to Eq.~\eqref{eqn:ResAddition}.


\subsection{Unit Variance Maintaining Layers}
\label{sec:architecture_variance}

As mentioned in Section~\ref{sec:architecture_data}, there are great benefits to keeping the input and output normalized within the neural network.
This is also the goal of batch normalization~\cite{BatchNormalization}, which explicitly normalizes the input into a specific layer in addition to introducing regularization. 
However, the regularization properties of batch normalization are too strong for the application at hand (see Section~\ref{sec:architecture_regularization}).
Therefore, an alternative, regularization-free method is defined which maintains a normalized throughput by ensuring that each layer conserves unit variance of its input.
Assuming that the input into a layer is zero centered and has unit variance,
the calculations performed in that layer can be modified such that the output remains normalized.
For instance, the sum~$z_\t{total}$ of $n$ independent random variables~$z_i$ has an increased variance
given by:
\begin{equation}
  s^2_\t{total} = \sum_i^n s^2_i,
  \label{eqn:VarianceAddition}
\end{equation}
where~$s^2_i$ is the variance of input $z_i$. 
If the inputs~$z_i$ have unit variance~($s^2_i = 1$), Eq.~\eqref{eqn:VarianceAddition} reduces to 
$s_\t{total} = \sqrt{n}$.
To compensate for this increased variance of the output, the result of the sum can be divided by $s_\t{total}$ to obtain the normalized output:
\begin{equation}
  z_\t{total}' = \frac{z_\t{total}}{s_\t{total}} =  \frac{z_\t{total}}{\sqrt{n}}.
  \label{eqn:VarianceAddition_normalized}
\end{equation}
Wherever applicable, transformations such as Eq.~\eqref{eqn:VarianceAddition_normalized} are applied.
The assumption of normalized inputs is an idealization that does not necessarily hold in practice.
Nevertheless, the application of these compensation factors helps in maintaining the variance to a reasonable range for the application described in this paper.

In addition to the modifications described above, trainable parameters of the neural network are initialized in a way that the layers initially perform an approximate identity operation.
This initialization scheme in combination with unit variance maintaining layers and the normalization of input and labels (see Section~\ref{sec:architecture_data}) greatly speeds up and facilitates the training procedure.
As a result, the initialized but untrained network predicts each label with a
probability proportional to its frequency in the training data.

\subsection{Regularization}
\label{sec:architecture_regularization}

In contrast to typical applications of CNNs in the domain of image recognition, 
applications in IceCube are not limited by the training dataset size.
The extremely large amount of available training data (on the order of hundreds of millions of events)
in combination with the rather small network capacity results in a reduced necessity for regularization.
A regularization technique called dropout~\cite{dropout}, which randomly drops input nodes in the training phase, is used in early stages of training and gradually reduced during the training procedure.
The last training steps are nearly performed regularization-free.
To ensure generalization, systematic variations of the baseline simulation are included in the training, validation, and test datasets.
The trained model is also tested for various modifications as discussed in Section~\ref{sec:systematics_systematics}.
Additionally, entire DOMs are dropped randomly during training via Dropout to ensure robust predictions and to emulate experimental data, where DOMs may drop
and be excluded for a certain data-taking period. 
This forces the neural network not to rely too much on the presence or values of a single DOM.


\subsection{Network Architecture}
\label{sec:architecture_architecture}

IceCube's data is split up and transformed to an orthogonal grid as described in Section~\ref{sec:data} and illustrated in Fig.~\ref{fig:IceCubeDataInput}.
The input data for each sub-array is sequenced into a series of convolutional layers as shown in Fig.~\ref{fig:architecture}.
A three dimensional convolution over all spatial coordinates is performed for the data of the main IceCube array, while a single convolution over the z-axis is used for the DeepCore arrays.
The result of these convolutional layers is flattened, combined, and is then used in two small fully-connected networks.
One of these fully-connected networks is used to obtain the prediction for each label quantity, whereas the other estimates the uncertainty on each predicted quantity.
The uncertainty-estimating sub-network also obtains the prediction output as input.
A gradient-stop\footnote{\url{https://www.tensorflow.org/versions/r1.15/api_docs/python/tf/stop_gradient}} is applied to the input of the uncertainty-estimating network such that its optimization does not alter the trainable parameters of the remaining network.
Details of the chosen network architecture and the layers of each sub-network are provided in Tab.~\ref{tab:architecture} and~\ref{tab:architecture_output} in Appendix~\ref{sec:appendix}.

During training, the dropout rates~$d_i$ are gradually decreased as defined in Section~\ref{sec:training_procedure}.
Entire DOMs of the input tensors are dropped with a rate of~$d_0$.
The convolutional layers use a dropout rate of~$d_1$ and a dropout rate of~$d_2$ is applied to the combined and flattened layer after the convolutional layers.
The values of the dropout rates~$d_i$ are provided in Tab.~\ref{tab:training_steps}.
An exponential linear activation function (elu)~\cite{ELUActivationFunction} is used for all layers except for the output layers.
The prediction sub-network does not use an activation function in its output layer, whereas the uncertainty sub-network uses the absolute value function (abs), enforcing non-negative values.

The network is set up to solve a multi-label regression task, although it can easily be altered to solve a classification task instead.
Its main focus is on the directional reconstruction of the primary particle and the deposited energy in the detector.
In principle, the directional reconstruction can be learned directly via the zenith and azimuth angle.
However, due to the azimuthal periodicity, it is beneficial to decompose the angles into the three components of the unit direction vector and to emphasize these during training.
The angles zenith~$\Theta$ and azimuth~$\Phi$ are therefore not directly estimated by the neural network, but are calculated from the normalized direction vector~$\vec{d}=(d_x, d_y, d_z)^T$ according to
\begin{align}
  \Theta &= \cos^{-1}\left(-d_z\right) \\
  \Phi &= \tan^{-1}\left(\frac{-d_y}{-d_x}\right) \pmod{2\pi}.
\end{align}
The uncertainty on the direction vector components~($\sigma_{d_x}$, $\sigma_{d_y}$, $\sigma_{d_z}$) and the
angles $\sigma_{\Theta}$ and~$\sigma_{\Phi}$ are estimated directly by the neural network.
For the sake of simplicity, correlations are not taken into account in this work.
They may be included by adding additional output nodes to the neural network for the off-diagonal elements of the covariance matrix in addition to changing the
loss function in Eq.~\eqref{eqn:GLLoss} to a multivariate Gaussian. 


%
%
%
%

\section{Training Data and Procedure}
\label{sec:training}

The architecture described in the previous section defines a multi-label regression task.
Training multiple labels at once can be challenging due to individual labels dominating the loss. 
Therefore, a weighted multi-label loss function is introduced.
In addition, the training data and procedure is described in this section.

\subsection{Training Datasets}
\label{sec:training_datasets}

The neural network is trained and evaluated on neutrino Monte Carlo (MC) simulations. 
Additional checks are performed on experimental data to validate the applicability of the MC simulations (see Section~\ref{sec:systematics}).
An event selection is performed on events that trigger the detector in order to reduce the atmospheric background and to improve the event quality. 
In this paper, the event selection described in Ref.~\cite{CascadePaperMike7yrs} is used, which aims to select cascade-like events (charged-current $\nu_e$ and $\bar\nu_e$ as well as neutral-current interactions of all neutrino types).

To investigate effects of systematic uncertainties in the MC simulation, 
datasets are simulated with discrete variations of key parameters, such as the scattering and absorption coefficients of the glacial ice.
These parameters affect the propagation of light in the detector medium and are a major source of systematic uncertainty.
The neural network is trained on the baseline MC dataset in addition to its systematic variations.
These systematic datasets are included in the early stages of training to become more robust towards these uncertainties.
The combined baseline and systematic dataset is split into three parts to obtain a training, validation, and test set.
The resulting training set has about six million events.

In order to efficiently train the neural network, loading, preprocessing, and sequencing the data into the network have to be highly optimized.
The chosen simulation dataset (about $\SI{10}{TB}$ of disk space) is preprocessed, compressed and saved to \textsf{hdf5} files.
During training, reading and decompression of the \textsf{hdf5} files is handled in parallel on multiple CPUs.
Once read and decompressed, the input tensors and labels of these events are combined from the multiple processes and enqued in a single shared queue.
An additional process dequeues the elements and produces batches of a given size, which are then sequenced into the neural network.

Due to the high dimensionality of the input and the restrictions given by the computing architecture, only a limited amount of files can be loaded into memory at once.
In order to reduce the amount of times files need to be read and decompressed, a given number of $m$ files are loaded at once.
Of the loaded and decompressed events, batches are randomly drawn until a specified amount of $k$ iterations through the data in memory (epochs) are completed.
Afterwards, the next $m$ files are loaded into memory.
This allows for the more efficient use of CPU resources, but some care must be taken to ensure
variability between consecutive training batches.
The values of $k$ and $m$ therefore depend on the batch size~$b$ and the number of events per file~$f$.
They are chosen such that the number of loaded events~$M = \sum_i^m f_i$ fulfills the relation~$M > \sqrt{k} \cdot b$.
The choice of the factor $\sqrt{k}$ as opposed to $k$ allows to reuse events in memory.
The larger the pool of loaded events~$M$ is with respect to the batch size, the higher is the number
of allowed epochs through the data in memory.

\subsection{Multi-label Regression}
\label{sec:training_multilabel}

The reconstruction method described in this paper is a multi-label regression task.
As such the neural network is trained on multiple labels simultaneously.
Due to the varying reconstruction difficulty of each label, 
their contribution to the loss function may be on different scales.
Note that the relative contributions of each label to the loss function define the contributions to the gradients used in gradient-descent during training.
Labels that are inherently hard to predict can dominate the training process,
preventing the accurate prediction of others.
To ensure that each label contributes equally to the loss function or by a predefined importance, 
an adaptive weighting method is proposed.

An importance vector~$\vec{C}$ is introduced which can be used to distribute weights to individual labels.
The importance vector has an entry for each label where $\left(\vec{C}\right)_k$ is the weight associated to the $k$-th label.
The loss~$\left( \text{L} \right)_{b,k}$ for an event~$b$ in a batch of $B$ events and label~$k$ can then be weighted by
\begin{equation}
  \left( \text{L}_w \right)_{b,k} = \left(\vec{C}\right)_k \cdot \left( \text{L} \right)_{b,k}.
  \label{eqn:WeightedLoss}
\end{equation}
%
Optionally, the events can additionally be weighted to reflect their
expected frequency in experimental data.
This changes the importance weighted loss~$\left( \text{L}_w \right)_{b,k}$ from Eq.~\eqref{eqn:WeightedLoss} to
\begin{equation}
  \left( \text{L}_w \right)_{b,k} = \left(\vec{W}\right)_b \cdot \left(\vec{C}\right)_k \cdot \left( \text{L} \right)_{b,k},
  \label{eqn:EventWeightedLoss}
\end{equation}
%
where $\vec{W}$ are the weights defining the expected frequency in experimental data for the $B$ events in a mini batch.
In this work, Eq.~\eqref{eqn:WeightedLoss} is used.
The overall scalar loss~$\text{L}_{\text{scalar}}$ is computed via
\begin{equation}
  \text{L}_{\text{scalar}} =\frac{1}{B} \sum\limits_{k=0}^K \sum\limits_{b=0}^B \left( \text{L}_w \right)_{b,k},
  \label{eqn:Loss}
\end{equation}
where $K$ is the number of labels and $B$ is the number of events in a batch.
The importance vector~$\vec{C}$ is updated every $N$ optimization steps to ensure that each label contributes to the loss function according to the importance assigned to it.
This compensates for the fact that different labels are trained at different speeds.
The importance vector is updated via:
\begin{equation}
  {\left( {\vec{C}'}\right)}_k = {\left( \vec{C_0}\right)}_k \cdot \max{ \left[ 1, { \left(\overrightarrow{\langle \t{MSE}\rangle}\right)_k }^{-\frac{1}{2}}\right]} ,
  \label{eqn:UpdateImportanceVector}
\end{equation}
where $\vec{C_0}$ is the unmodified, original importance vector and $ \overrightarrow{\langle \t{MSE}\rangle}$ is given by
\begin{equation}
  \overrightarrow{\langle \t{MSE}\rangle} = \frac{1}{N} \sum\limits_{n=0}^{N} \overrightarrow{\t{MSE}}_n
  \label{eqn:AverageMSE}
\end{equation}
with the mean squared error~$\overrightarrow{\t{MSE}}_n$ over each normalized label for the $n$-th mini batch.
A rolling average can also be chosen as an alternative to the update rule from Eq.~\eqref{eqn:UpdateImportanceVector}.

\subsection{Loss Functions}
\label{sec:training_loss}

During the first training steps, both the predictions and uncertainty estimates are trained
by loss functions employing mean squared errors (MSE) due to their robustness.
The loss~$\left( \text{L} \right)_{b,k}$ from Eq.~\eqref{eqn:WeightedLoss} is therefore given by
\begin{equation}
  \left( \text{L} \right)_{b,k} = \underbrace{\left(\left(Y''_\t{true} - Y''_\t{pred}\right)_{b, k}\right)^2}_{\left( \text{L}_\text{pred}\right)_{b, k}} + 
  \underbrace{\left[ \left(Y''_\t{unc}\right)_{b, k} - \texttt{gradient\_stop}\left(  \left|\left(Y''_\t{true} - Y''_\t{pred}\right)_{b, k} \right|\right) \right]^2}_{\left(\text{L}_\text{unc}\right)_{b, k}}
  \label{eqn:MSELoss}
\end{equation}
where L, $Y''_\t{true}$, $Y''_\t{pred}$, and $Y''_\t{unc}$ are $B\times K$-matrices with the number of labels~$K$ and the number of events~$B$ in a batch. 
The true normalized labels are given by $Y''_\t{true}$ (see Eq.~\eqref{eqn:normalization}).
$Y''_\t{pred}$ is the normalized network prediction and $Y''_\t{unc}$ is the normalized output of the uncertainty estimating sub-network.
The function $\texttt{gradient\_stop}(x)$ is an identity operation that effectively treats $x$ as a constant during backpropagation.
Optimization of $\text{L}_\text{unc}$ therefore only affects the weights of the uncertainty estimating sub-network due to the applied gradient-stop. 
Gradients are not propagated through to the main neural network architecture.
The gradient-stop is introduced in order to stabilize the initial training iterations
and to allow for independent optimization of the two sub-networks.

In later training stages, once the loss starts to converge, 
i.e. the improvement in loss per training step starts to flatten out, 
the neural network's reconstruction and uncertainty estimation is robust enough for the application of the more sensitive Gaussian Likelihood (GL).
In this case, the loss changes to
\begin{equation}
  \left( \text{L} \right)_{b,k} = 2\cdot \ln{\left( \left(Y''_\text{unc}\right)_{b, k} \right)} + \left( \frac{\left(Y''_\t{true} - Y''_\t{pred}\right)_{b, k}}{\left(Y''_\text{unc}\right)_{b, k}} \right)^2.
  \label{eqn:GLLoss}
\end{equation}
When employing Eq.~\eqref{eqn:GLLoss} as a loss function, 
the presented network architecture may be interpreted as a 
mixture density network~\cite{MixtureDensityNetworks}
with a single Gaussian kernel.

\subsection{Training Procedure}
\label{sec:training_procedure}

The loss function~$\text{L}_{\text{scalar}}$ from Eq.~\eqref{eqn:Loss} is minimized with the ADAM-optimizer~\cite{ADAMOptimizer} within the \textsf{TensorFlow} framework~\cite{tensorflow2015-whitepaper}.
The general learning scheme is to start training with a high dropout rate, forcing the network to learn robust features.
Over time the learning and dropout rate are reduced.
During early stages of training, the neural network is also trained on systematic variations of the baseline dataset to promote robustness towards uncertainties in the MC simulation.
Training is performed in a process of multiple steps as depicted in Tab.~\ref{tab:training_steps} and takes about five days on an NVIDIA Tesla P40 GPU.
\section{Reconstruction Performance}
\label{sec:performance}

Although the CNN architecture is a versatile tool, emphasis is put on the performance of cascade-like events for the purpose of this paper.
The architecture can easily be used to reconstruct other event topologies or labels of interest.
Minor changes to the loss function and activation of the last layer can be applied to use the architecture
for classification tasks.
In the following sections, the performance of the described CNN architecture is compared to the current standard cascade reconstruction method in IceCube (see Section~\ref{sec:cascades_reco} for details).
Results shown in this paper are obtained for the cascade event selection described in Ref.~\cite{CascadePaperMike7yrs}, 
which aims to select high-energy events in the TeV range and above resulting from charged-current $\nu_e$ and $\bar\nu_e$ as well as neutral-current interactions of all neutrino types.
Wherever applicable, results shown for the CNN-based method are denoted by a {\includegraphics[scale=0.053]{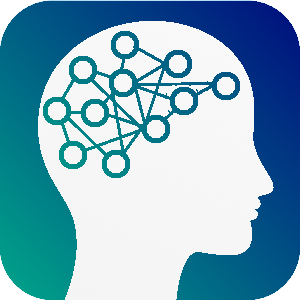}} icon and results for the standard (approximate) likelihood-based reconstruction are denoted by a $\mathcal{L}$ symbol.

\subsection{Angular Resolution}
\label{sec:performance_angle}

\begin{figure}
  \begin{minipage}[t]{0.63\textwidth}
    \vspace{0pt}
    \includegraphics[width=\textwidth, keepaspectratio]{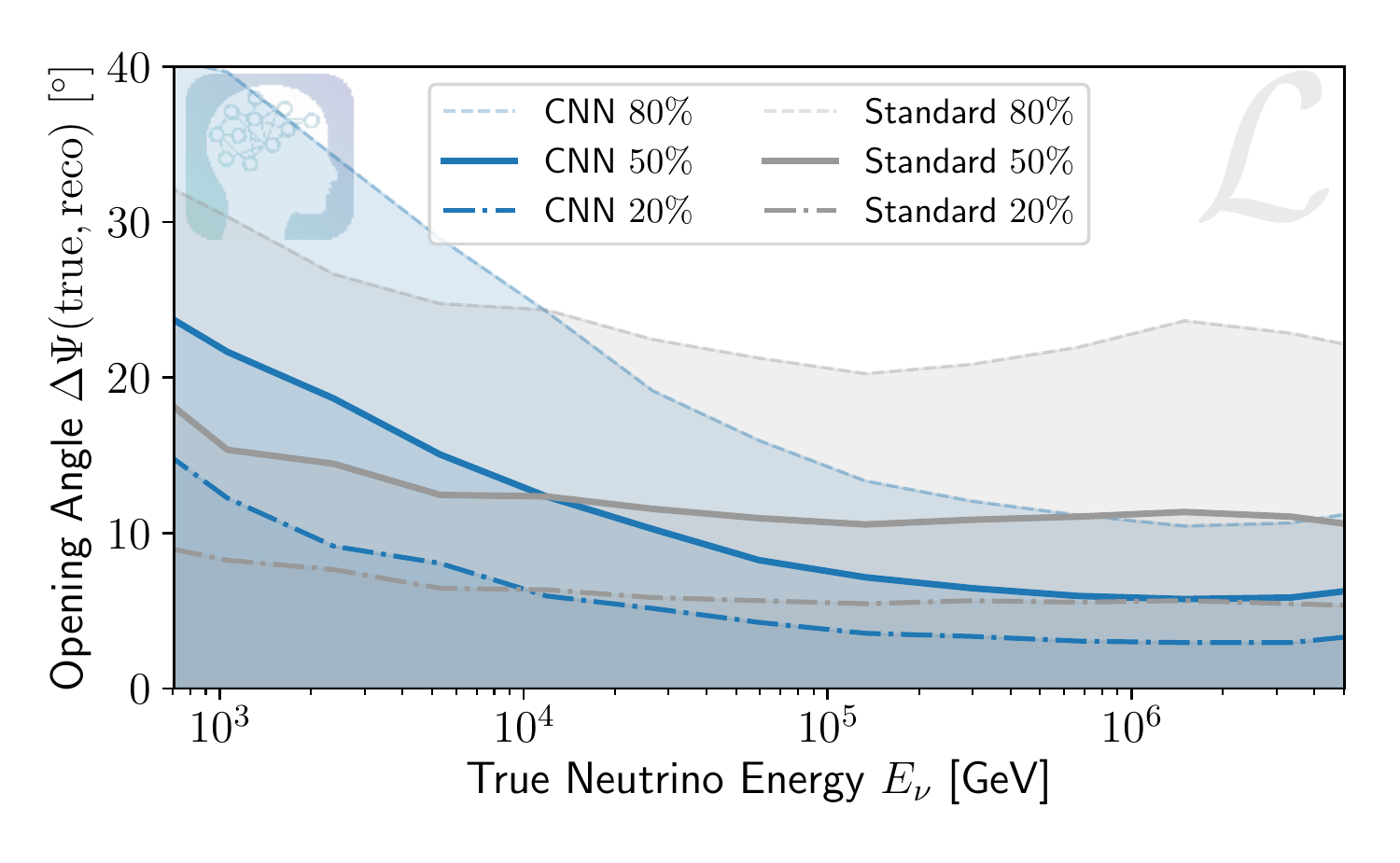}
  \end{minipage}\hfill
  \begin{minipage}[t]{0.36\textwidth}
    \vspace{10pt}
    \caption{The angular resolution is shown as a function of neutrino energy for the standard reconstruction method and the newly developed CNN-based method. The shaded area and lines denote the 20\%, 50\%, and 80\% quantiles. At higher energies, the resolution can be improved by up to $\SI{50}{\percent}$. Systematic uncertainties are not included.} 
    \label{fig:angular_resolution}
  \end{minipage}
\end{figure}

Because the statistical power of searches for neutrino sources depends upon the ability to localize the origin of neutrino candidates, improving the precision of directional reconstruction is of great importance.
In order to evaluate the precision of the directional reconstruction, the metric angular resolution is introduced, which is defined as the median opening angle between the true and reconstructed direction.
The relative contribution of astrophysical neutrinos steadily increases with energy.
At lower energies, the neutrino candidates in the event sample are dominated by atmospheric background events.
The CNN-based method described in this paper is able to improve the angular resolution of neutrinos above $\SI{10}{TeV}$.
Fig.~\ref{fig:angular_resolution} compares the angular resolution of the standard reconstruction method and the newly developed CNN-based method. 
At higher energies, the resolution can be improved by up to $\SI{50}{\percent}$.
Using the CNN for reconstruction affords improved sensitivity at higher energies compared to standard reconstruction methods.
In combination with additional years of data, this results in an improved point source sensitivity up to a factor of four in the southern sky~\cite{CascadePaperMike7yrs} compared to previous work based on standard reconstruction methods~\cite{CascadePaper2yrs}.

%

\subsection{Energy Resolution}
\label{sec:performance_energy}

The primary neutrino energy cannot be directly measured in IceCube,
but it can be inferred from the deposited energy in the detector.
For charged-current cascade-like events, the deposited energy is an excellent proxy for the true neutrino energy as detailed in Ref.~\cite{EnergyReconstruction}.
Neutral-current events are more ambiguous due to large fluctuations in
the fraction of the neutrino energy transferred to the target nucleus.
Further smaller uncertainties arise from fluctuations in the light yield of hadronic cascades.
Therefore, cascade reconstruction methods reconstruct the electro-magnetic (EM) equivalent deposited energy.
This is the energy of an EM cascade that produces the same number of
Cherenkov photons as the observed shower.

\begin{figure}
  \begin{center}
    \includegraphics[width=\textwidth, keepaspectratio]{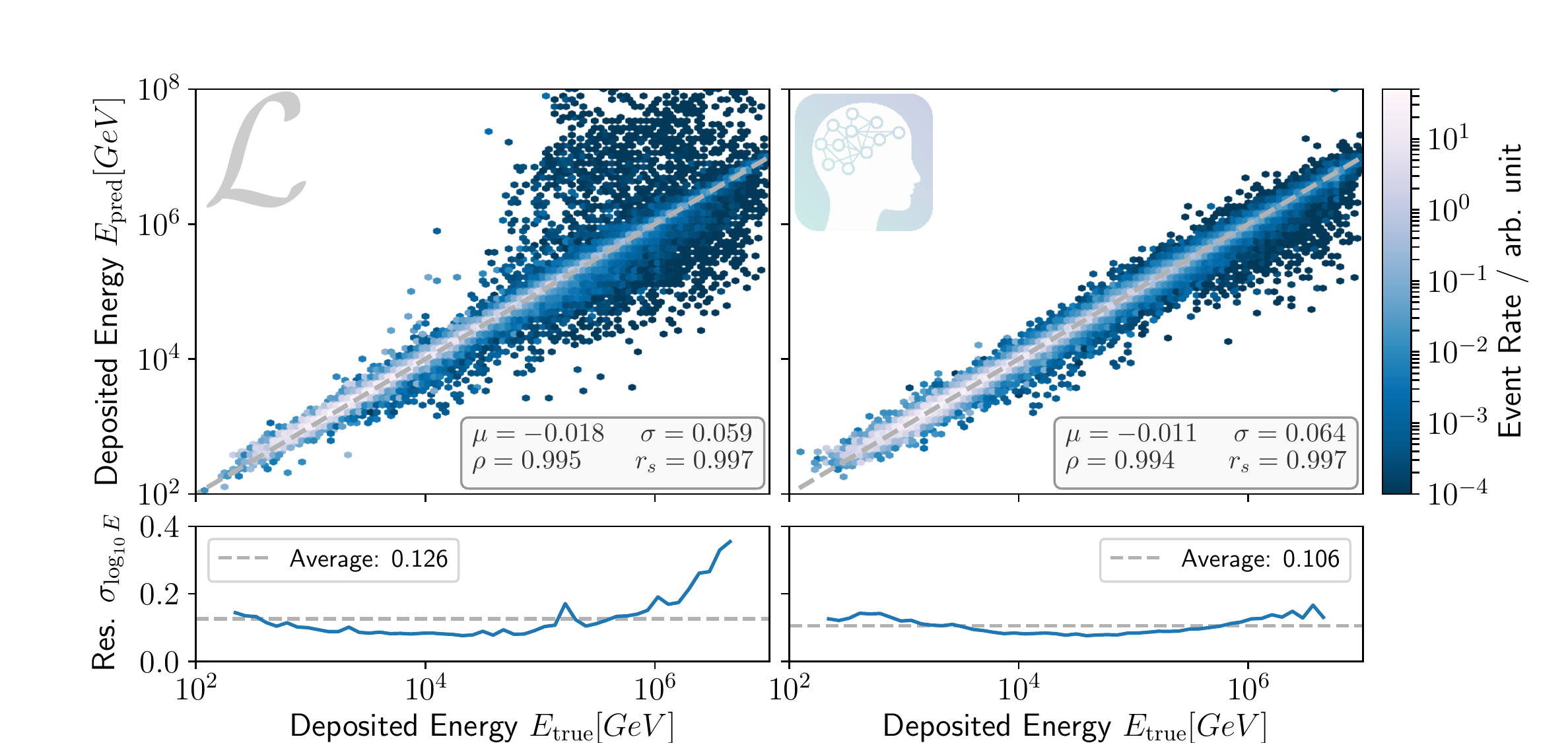}
    \caption{Correlation plots between the true and reconstructed deposited energy are shown for the standard reconstruction (left) and CNN (right). Mean~$\mu$ and standard deviation~$\sigma$ of the residuals $y_\text{pred} - y_\text{true}$ are determined in $\text{log}_{10}$-space as well as the Pearson~$\rho$ and Spearman~$r_s$ correlation coefficients. The energy resolution in the bottom panel is calculated according to~\cite{EnergyReconstruction}.}
    \label{fig:energy_resolution}
  \end{center}
\end{figure}
The neural network is trained on this EM equivalent energy.
For hadronic cascades, the average expected EM equivalent energy is chosen.
Overall, the energy resolution of the standard reconstruction and the CNN-based method are comparable
as shown in Fig.~\ref{fig:energy_resolution}.
There are slight biases visible in the CNN estimate at lowest and highest energies.
These are artifacts induced by the boundaries of the training dataset.
If necessary, the CNN estimate can be calibrated to remove this bias.
In comparison to the standard reconstruction, the predictions of the CNN are more robust and have less outliers.
There is a considerable smearing visible in the correlation plot for the standard reconstruction method at higher energies.

\subsection{Uncertainty Estimation}
\label{sec:performance_uncertainty}

\begin{figure}[!tbp]
  \centering
  \begin{minipage}[t]{0.41\textwidth}
    \vspace{0pt}
    \includegraphics[width=\textwidth, keepaspectratio]{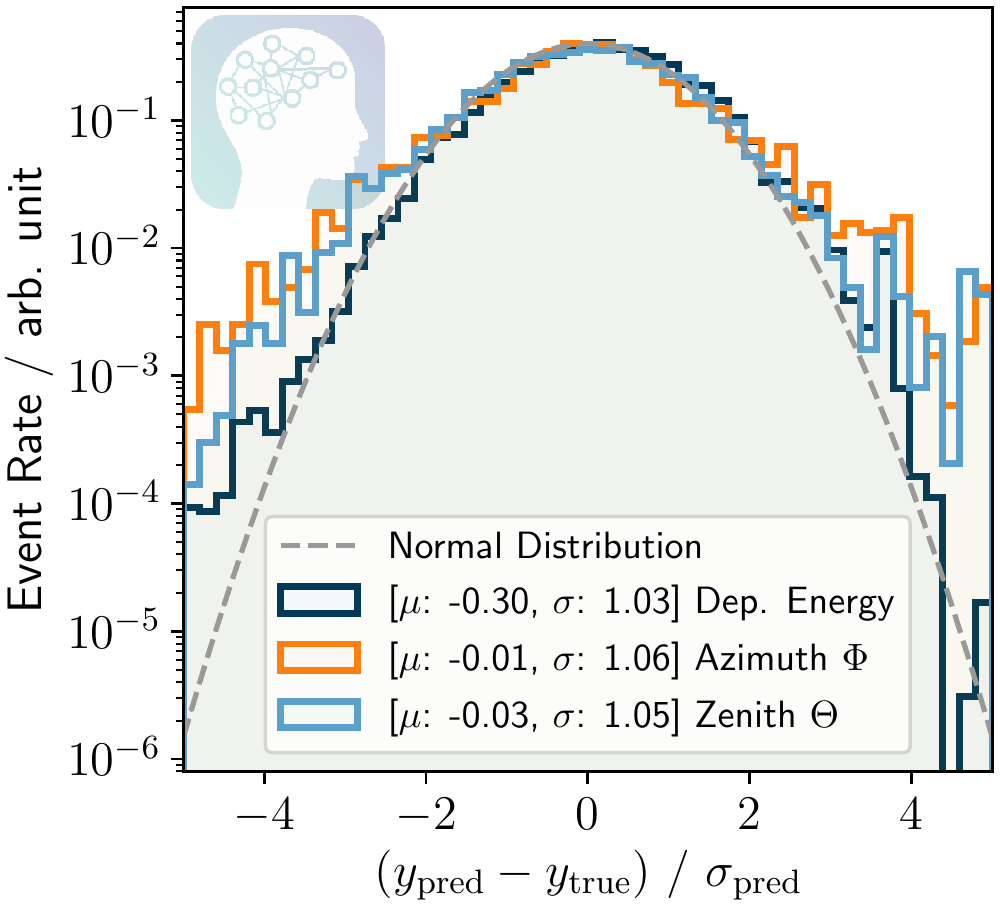}
    \caption{The pull distributions are shown for the reconstructed quantities in comparison to the normal distribution which indicates a good uncertainty estimator. These match well, apart from slight deviations in the tails of the distribution.}
    \label{fig:pull}
  \end{minipage}
  \hfill
  \begin{minipage}[t]{0.55\textwidth}
    \vspace{0pt}
    \includegraphics[width=\textwidth, keepaspectratio]{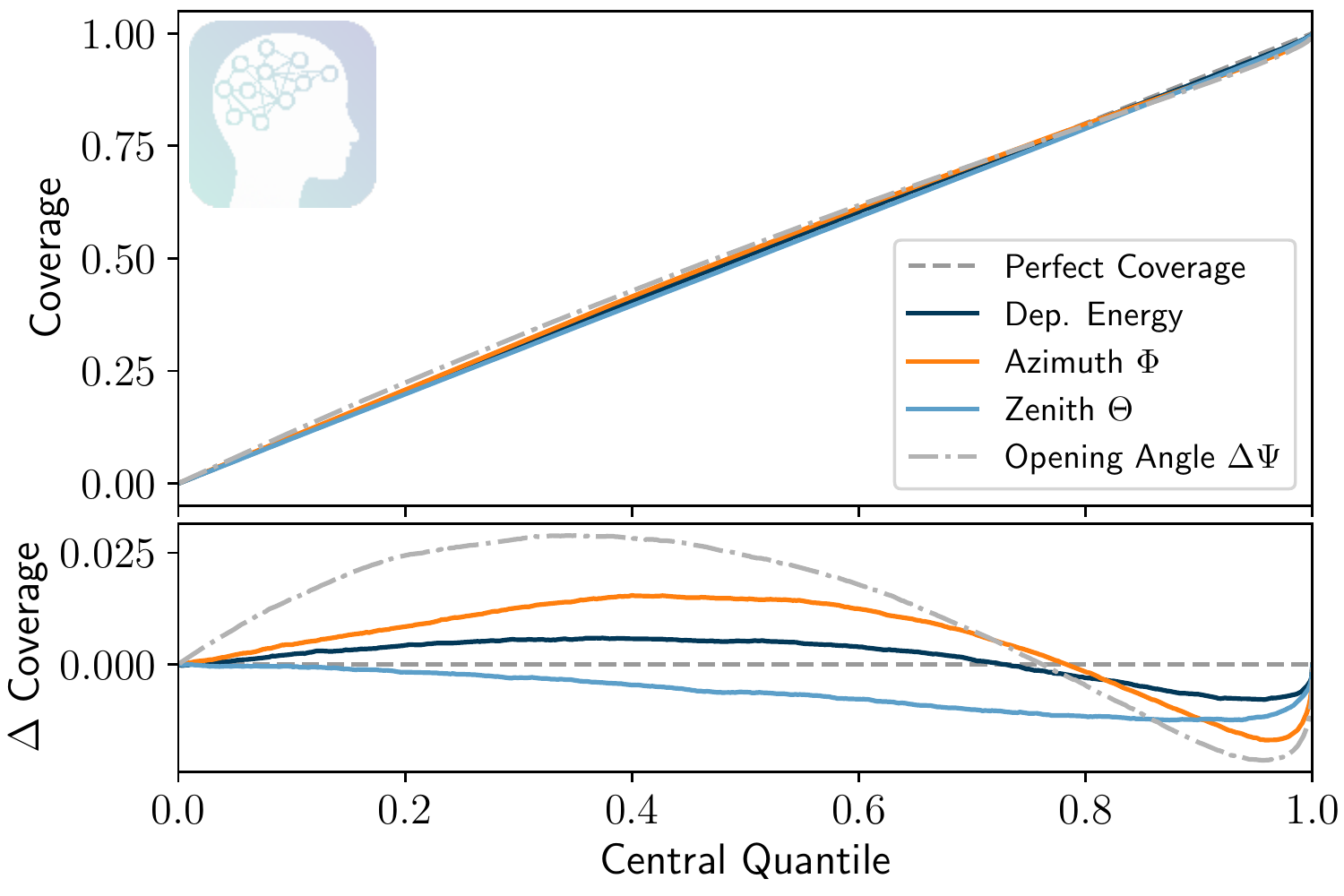}
    \caption{The coverage, i.e. the number of events that fall in a certain quantile, is shown as a function of the computed central quantile based on the uncertainty estimate. Perfect coverage is obtained on the dashed diagonal. The bottom panel shows the difference in coverage of each reconstructed quantity compared to the perfect one-to-one relation.}
    \label{fig:coverage}
  \end{minipage}
\end{figure}

For application of a reconstruction method in analyses, it is necessary to have a measure on the uncertainty
of a reconstructed quantity.
As described in Section~\ref{sec:architecture_architecture} and~\ref{sec:training_loss}, the model architecture estimates a per-event
uncertainty on each of its reconstructed quantities via an assumed Gaussian likelihood.
This works well, if the residuals $\Delta y = y_\text{pred} - y_\text{true}$ follow a Gaussian distribution with a given event dependent standard deviation. 
In many cases, the assumed Gaussian likelihood is a good fit as illustrated in Fig.~\ref{fig:pull} and~\ref{fig:coverage}
which show the pull distribution and the coverage of the estimated uncertainties for the ensemble of events, respectively.
%

A correct uncertainty estimator for Gaussian residuals will produce a pull distribution that converges to a normal distribution in the limit of infinite statistics. 
The pull distributions shown in Fig.~\ref{fig:pull} are reasonably well described by a normal distribution, indicating correct uncertainty estimates.
There are slight deviations visible in the far tail of the distribution which are caused by rare outlier events.
These outliers might be reduced through additional training iterations or by a biased sampling strategy during training that enriches the training batches with mis-reconstructed events.

Another way of validating the estimated uncertainty is by checking its coverage.
Based on the assumption of a Gaussian distribution with a width as estimated by the neural network, the number of events can be calculated that should lie within a certain range around the prediction.
This number can then be compared to how many events actually fall inside.
For an accurate uncertainty measure, this should result in a one-to-one relation described by the dashed diagonal in Fig.~\ref{fig:coverage}.
The uncertainty estimates on all reconstructed quantities achieve an excellent coverage on the ensemble
which deviates only by a few percent from the case of perfect coverage.

As detailed in Section~\ref{sec:architecture_architecture}, 
the CNN estimates the uncertainty on each reconstructed quantity independently
while disregarding potential correlations.
In order to estimate the angular uncertainty, which is more closely related to the angular
resolution defined as the median opening angle~$\Delta\Psi$
between reconstructed and true direction, either the uncertainties on the direction vector
components~($\sigma_{d_x}$, $\sigma_{d_y}$, $\sigma_{d_z}$) or the uncertainties on the 
angles $\sigma_{\Theta}$ and~$\sigma_{\Phi}$ have to be combined.
Here, the following prescription is chosen:
\begin{equation}
  \sigma_{\Delta\Psi} = \sqrt{\frac{\sigma_{\Theta}^2 + \left(\sigma_{\Phi}\cdot\sin{\Theta_{\text{CNN}}}\right)^2}{2}}
  \label{eqn:circular_err}
\end{equation}
to obtain a simplified ``circularized'' 
uncertainty estimate~$\sigma_{\Delta\Psi}$ for the angular resolution.
$\Theta_{\text{CNN}}$ in Eq.~\eqref{eqn:circular_err} is the reconstructed zenith angle by the CNN.
This ``circularized'' uncertainty estimate assumes that the uncertainty on the direction
reconstruction may be approximated by a circle in zenith-azimuth-space.
If correlations exist or if the uncertainties in reconstructed zenith and azimuth angles are
on different size scales, this simplified assumption will lead to under or over-coverage.
Fig.~\ref{fig:coverage} shows that the ``circularized'' uncertainty estimate~$\sigma_{\Delta\Psi}$ 
obtains correct coverage on average for the ensemble.
The coverage for $\sigma_{\Delta\Psi}$ is computed assuming a Rayleigh distribution.

Contrary to maximum likelihood estimation which can evaluate the local curvature of the likelihood to estimate per-event uncertainties, the neural network is a data driven approach.
As such it relies on proper coverage of the phase space through the training data.
Certain areas in the phase space that are less populated by training data, might therefore lead to
biases and result in over- or under-coverage of the estimated uncertainty.
Fig.~\ref{fig:pull} and~\ref{fig:coverage} show that the CNN-based uncertainties can obtain correct
coverage on average for the ensemble of events.

The results shown here are only for statistical uncertainties.
%
Known systematic uncertainties can, however, be included by adding training examples from systematic datasets throughout the entire training process.
To obtain proper coverage, the training events must be sampled or weighted according to the assumed priors on the systematics.
Ideally, the training events are sampled from a continous distribution of systematic variations.
This can be achieved by application of the SnowStorm method described in Ref.~\cite{SnowStorm}.
The CNN presented here only includes events from the systematic dataset in early stages of the training to promote robustness.
Later training steps are only performed on the baseline dataset.
The neural network is therefore trained to estimate the statistical uncertainty, whereas systematic uncertainties are not included in the prediction.

\subsection{Runtime}
\label{sec:performance_runtime}

\begin{figure}
  \begin{minipage}[t]{0.55\textwidth}
    \vspace{0pt}
    \includegraphics[width=\textwidth, keepaspectratio]{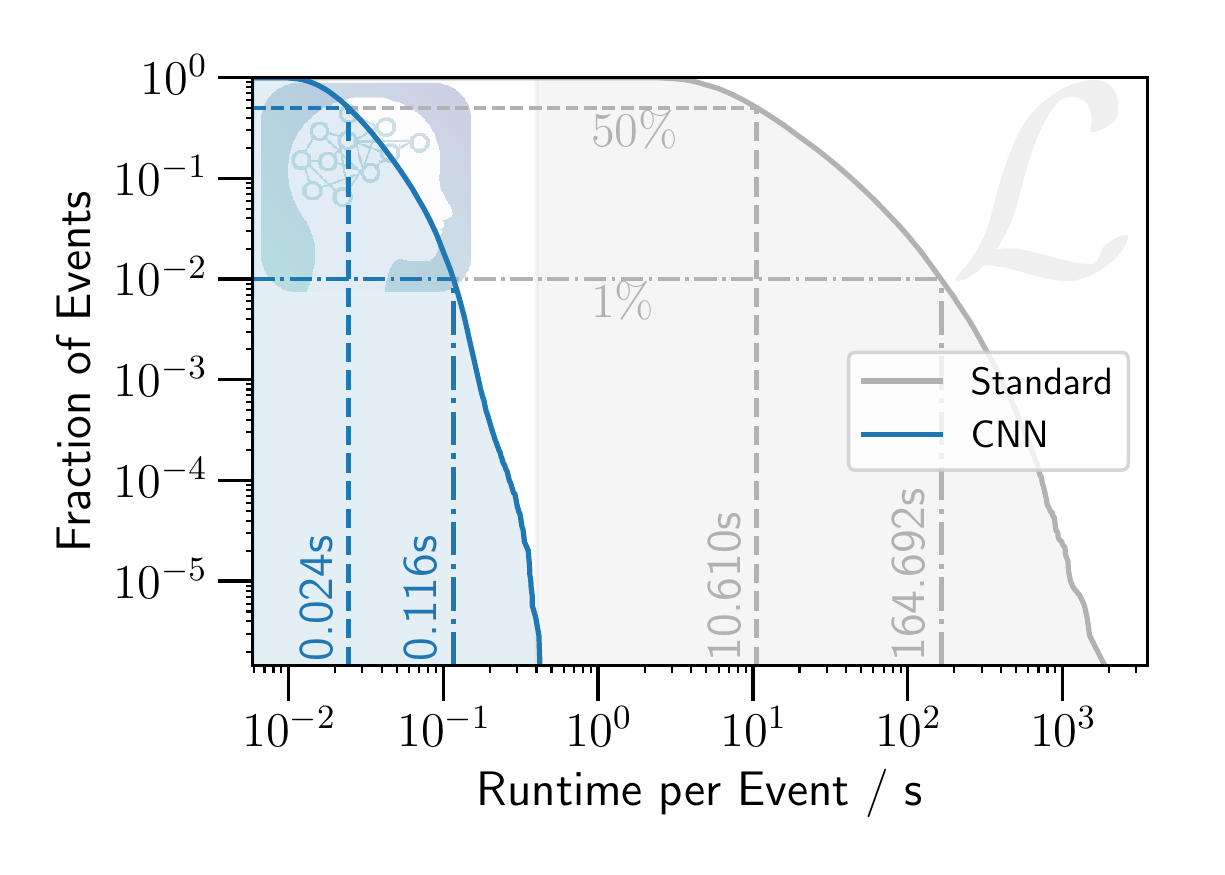}
  \end{minipage}\hfill
  \begin{minipage}[t]{0.44\textwidth}
    \vspace{10pt}
    \caption{
    Per-event runtimes are shown as a survival function of the fraction of events exceeding a specified runtime.
    In contrast to the standard reconstruction, the CNN-based method is able to run on multiple cores as well as on a GPU. 
    Runtimes shown for the CNN are for application on a GPU. The CNN outperforms the standard method by 2-3 orders of magnitude. 
    } 
  \label{fig:runtime}
  \end{minipage}
\end{figure}

A key advantage and motivation for the CNN-based reconstruction method is the desired application within IceCube's real-time system as mentioned in Section~\ref{sec:intro}.
Strict hardware requirements adhere to reconstructions run at the South Pole due to limited resources on-site.
In addition to the requirement of a fast per-event runtime, the reconstruction methods should complete in a foreseeable and ideally constant time in order to prevent pileup.
For most advanced likelihood-based reconstruction methods, this can pose a considerable challenge.
The maximization of the likelihood can be prohibitively expensive and, more importantly, its runtime may vary over many orders of magnitude 
as indicated by the width of the shaded area in Fig.~\ref{fig:runtime}.

%

Figure~\ref{fig:runtime} compares the per-event runtime between the CNN-based method (blue) and the standard reconstruction (gray).
The standard reconstruction method which depends on the maximization of an approximative likelihood has a wide distribution of runtimes that spans over nearly four orders of magnitude.
With a median runtime of about $\SI{10}{s}$, the standard reconstruction can be applied offline, but it is too slow to use on-site.
In comparison, the CNN-based method finishes in less than $\SI{24}{ms}$ for half of all reconstructed events.

The neural network performs a fixed amount of computations defined by the chosen architecture and independent of the event parameters.
Therefore, the time necessary for one forward pass
is nearly constant and takes about~$\SI{6}{ms}$ on an NVIDIA GTX 980.
The use of the CNN with a CPU increases the runtime compared to a GPU by a factor of \textasciitilde 60.
The bottleneck of the CNN-based method, if run on a GPU, is the data preparation step which includes the calculation of the DOM summary statistics (Section~\ref{sec:data_dom_inputs}).
This step also introduces a dependence on the amount of detected pulses in an event and is responsible for the widening of the runtime distribution shown in Fig.~\ref{fig:runtime}.
Although the runtime of the CNN-based method is fast enough, it can be further reduced through optimization of the input pipeline or by choosing less computing intensive input variables.
An end-to-end solution starting directly from the measured pulses as described in Section~\ref{sec:data_dom_inputs} may also decrease the runtime.
If run on a CPU, the bottleneck shifts from the data preparation step to the forward pass through the network.
In this case, the size of the network architecture can be decreased while only minimally affecting its performance.

\section{Robustness and Data/Simulation Agreement}
\label{sec:systematics}
 
 In the previous section, the performance of the CNN-based method was shown 
to offer comparable, and for some tasks, even improved performance relative to the
 standard cascade reconstruction method in IceCube.
 However, the performance plots shown are for the MC baseline dataset.
Limits in our knowledge of the detector and the underlying physics result in systematic uncertainties in the MC simulation.
Some of the sources of systematics are known, such as the scattering and absorption coefficients of the glacial ice.
The effect of these systematic uncertainties can be studied in dedicated simulations for which these parameters are varied.
More challenging are unknown systematics that cannot be simulated and explicitly tested.
A key quality parameter of a reconstruction method is therefore its data/MC agreement and the robustness of the method towards possible uncertainties in the MC simulation.
In the context of this paper, robustness of a reconstruction method is defined as its insensitivity
towards changes in the input data.
Note that a robust method does not necessarily have to be an accurate one.
In contrast, often a trade-off between reconstruction accuracy and robustness exists.
In the following sections, the data/MC agreement of the presented reconstruction method will be investigated and its robustness tested against potential mis-modeling in the simulation.

\subsection{Agreement for Baseline Simulation}
\label{sec:systematics_agreement}

A correct MC simulation will not contain differences to experimental data, 
although in practice, the simplified model in the simulation will not 
perfectly describe the data.
While a simulation cannot accurately describe all low level physics, it is important that the resulting high-level variables, from which physics results are derived, are well described.
Reconstruction methods produce such high-level variables and must therefore be robust towards low-level disagreements in the simulation.
The data/MC agreement of a given reconstruction method can be quantified by comparing distributions of analysis level parameters. 

\begin{figure}
  \begin{center}

    \includegraphics[width=\textwidth, keepaspectratio]{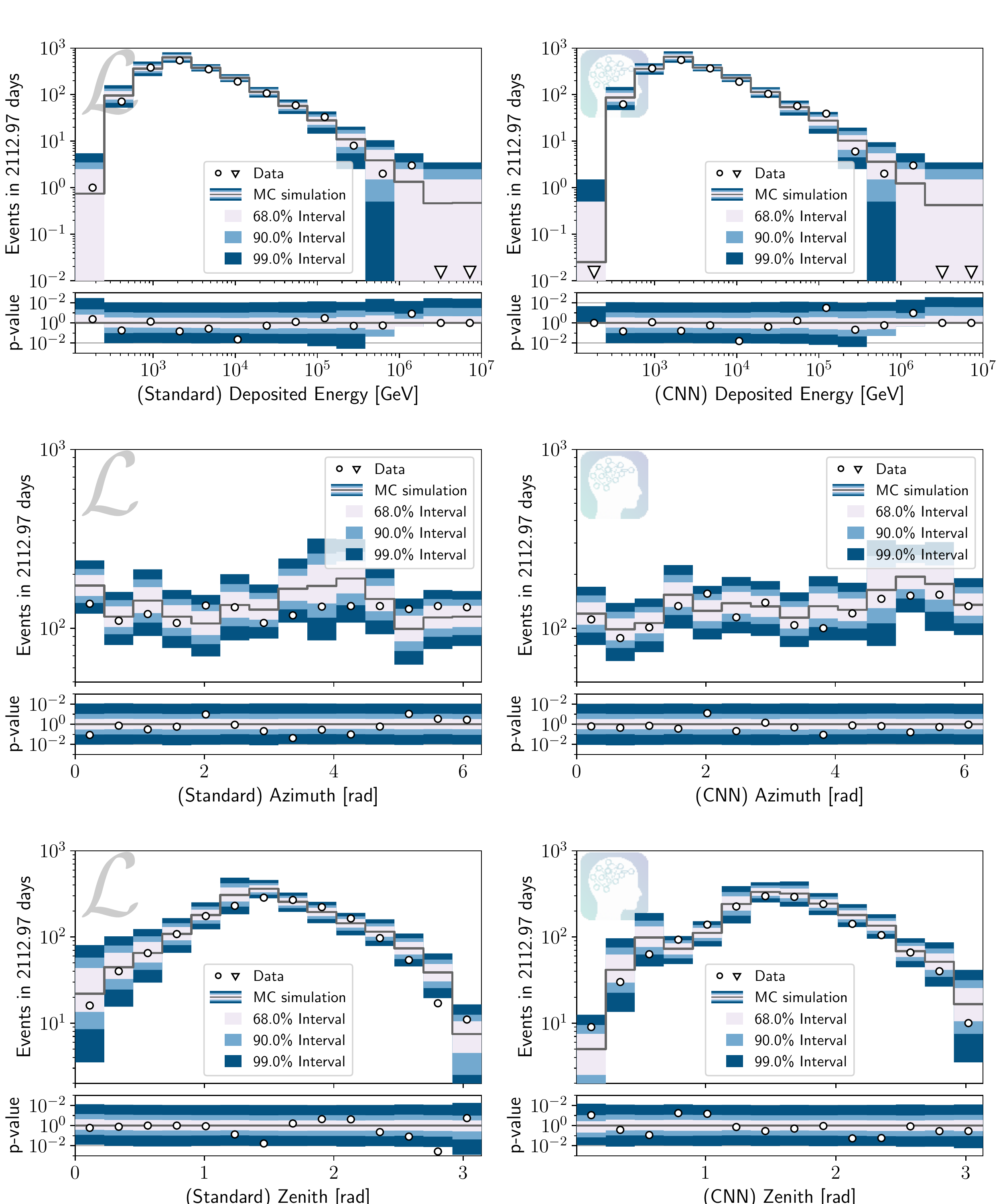}
    
    \caption{The simulated and measured distributions for the reconstructed quantities are shown. The left side shows the standard reconstruction and the right side shows the CNN-based method. Each plot is divided into two panels showing the number of events (top panel) and the significance of the fluctuations (bottom panel) in each bin. The significance calculation is based on the assumption of Poisson-distributed data and does not include systematic uncertainties.}
    \label{fig:data_mc}
  \end{center}
\end{figure}

The agreement between simulated and experimental data for the distributions of the reconstructed zenith, azimuth, and deposited energy is investigated in Fig.~\ref{fig:data_mc}.
On the left side, the distributions are compared for the standard reconstruction method and on the right for the CNN.
The distributions and their agreement is comparable between the two reconstruction methods.
Some smaller fluctuations exist, but overall, the simulation and the reconstructed quantities well describe the experimental data.

The comparisons of the one-dimensional distributions in Fig.~\ref{fig:data_mc} are blind to some classes of multivariate disagreements.
In principle, higher-dimensional distributions can be compared, but the data will become too sparse once the dimensionality reaches a certain level.
An alternative approach is to employ a multivariate classifier to distinguish data from simulation~\cite{DataMCMismatch, EventReweighting}.
A random forest (RF)~\cite{RandomForest} classifier from the \textsf{scikit-learn}~\cite{scikit-learn} package is trained in a 5-fold cross-validation.
To quantify the found mis-match, the area under curve (AUC) metric of the receiver-operating curve is used.
Ideally, the classifier should not be able to tell the difference between simulated and experimental data, i.e. the AUC should be around $0.5$.
If it is capable of separating the classes ($\text{AUC} > 0.5$), the feature importance vector of the random forest can be used to understand the origin of the disagreement.
The feature importances hint at variables that are important for the separation task and might therefore have a higher mis-match.

\begin{figure}
  \begin{minipage}[t]{0.49\textwidth}
    \vspace{0pt}
    \includegraphics[width=\textwidth, keepaspectratio]{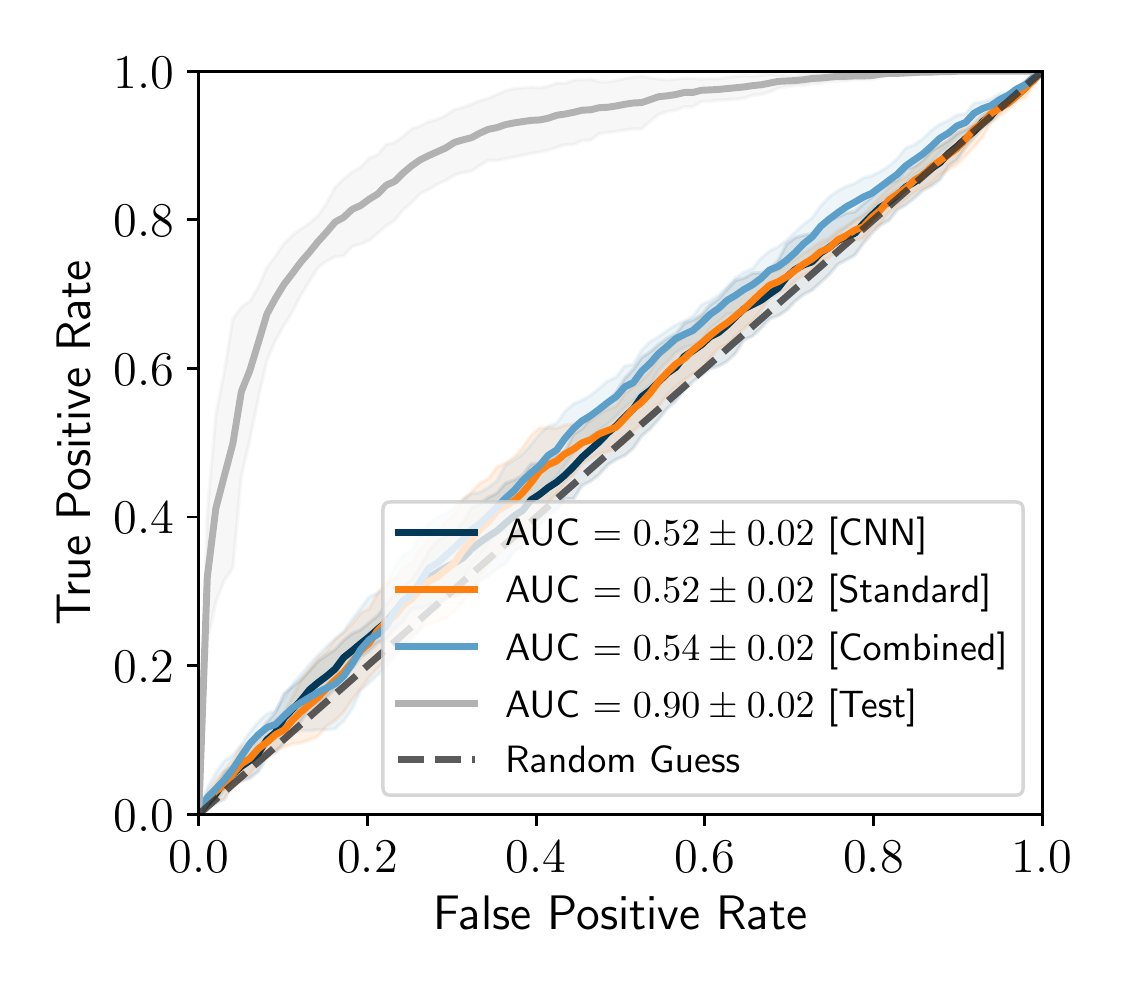}
  \end{minipage}\hfill
  \begin{minipage}[t]{0.49\textwidth}
    \vspace{10pt}
    \caption{A random forest (RF) is trained in a 5-fold cross-validation to distinguish data events from simulated ones on the basis of different sets of features. The receiver-operating curves are shown for each set of features. The area under this curve (AUC) is a metric for the classifier performance. A value of $1.0$ denotes a perfect classification while randomly assigning classes will result in an AUC of $0.5$. Events are not distinguishable on the basis of the reconstructed values for either of the reconstruction methods.} 
    \label{fig:roc_curve}
  \end{minipage}
\end{figure}

To test the applicability of this method, a RF is trained on the output variables of both reconstruction methods in addition to artificially created variables that contain a certain mis-match.
Variables $\text{Var}_{x, c}$ and $\text{Var}_{y, c}$ are added which are drawn from a bivariate normal distribution with a covariance of $+c$ for data events and $-c$ for simulated events.
The one-dimensional distributions of these variables for data and simulated events are indistinguishable by construction.
Only a multivariate approach is able to find the disagreement. 
The resulting receiver-operating curve is shown in Fig.~\ref{fig:roc_curve} labeled as "Test".
As expected, the RF is able to distinguish the events based on the added variables which contain mis-matches in correlated variables.
The resulting feature importances also correctly order the artificial features according to their mis-match as shown on the left of Tab.~\ref{tab:importances}.
%

%

\begin{table}
  \caption{Feature importances obtained from the RF, trained to distinguish simulated from experimental data, are shown for each of the four sets of features. The test case in which artificial features~$\text{Var}_{x/y, c}$ with varying degrees of mis-matches are added is shown on the left. An AUC score of \protect$0.90 \pm 0.02$ is achieved. 
The importances of the CNN-only (AUC = \protect$0.52 \pm 0.02$), standard reconstruction-only (AUC = \protect$0.52 \pm 0.02$), and combined feature set 
including outputs from both reconstructions
(AUC = \protect$0.54 \pm 0.02$) 
are shown on the right from top to bottom, respectively.}
  \label{tab:importances}
  \begin{center}
	\renewcommand*{\arraystretch}{1.04}
	\begin{tabular}{ll}
		\multicolumn{2}{c}{Test} \\
		\toprule
		Feature Name & Importance \\
		\midrule
		$\mathrm{Var}_{x, c=0.9}$ & $0.09\pm0.07$ \\
		$\mathrm{Var}_{y, c=0.9}$ & $0.08\pm0.07$ \\
		$\mathrm{Var}_{x, c=0.5}$ & $0.06\pm0.05$ \\
		$\mathrm{Var}_{y, c=0.5}$ & $0.05\pm0.04$ \\
		$\mathrm{Var}_{y, c=0.0}$ & $0.03\pm0.02$ \\
		CNN - Zenith & $0.03\pm0.02$ \\
		Standard - Zenith & $0.03\pm0.02$ \\
		CNN - Energy Unc. & $0.03\pm0.02$ \\
		Standard - Energy & $0.02\pm0.02$ \\
		CNN - Azimuth Unc. & $0.02\pm0.02$ \\
		CNN - Energy & $0.02\pm0.02$ \\
		$\mathrm{Var}_{x, c=0.0}$ & $0.02\pm0.02$ \\
		CNN - Azimuth & $0.02\pm0.02$ \\
		CNN - Zenith Unc. & $0.02\pm0.02$ \\
		Standard - Azimuth & $0.02\pm0.02$ \\
		\bottomrule
	\end{tabular}
	\qquad
	\renewcommand*{\arraystretch}{1.0}
	\begin{tabular}{ll}
		\multicolumn{2}{c}{Real} \\
		\toprule
		Feature Name & Importance \\
		\midrule
		CNN - Zenith & $0.18\pm0.17$ \\
		CNN - Energy & $0.17\pm0.16$ \\
		CNN - Azimuth & $0.17\pm0.16$ \\
		\midrule
		Standard - Zenith & $0.17\pm0.17$ \\
		Standard - Azimuth & $0.17\pm0.16$ \\
		Standard - Energy & $0.16\pm0.16$ \\
		\midrule
		Standard - Zenith & $0.07\pm0.06$ \\
		CNN - Energy Unc. & $0.06\pm0.06$ \\
		CNN - Zenith & $0.06\pm0.06$ \\
		CNN - Azimuth & $0.06\pm0.05$ \\
		Standard - Azimuth & $0.06\pm0.05$ \\
		CNN - Azimuth Unc. & $0.06\pm0.05$ \\
		CNN - Zenith Unc. & $0.06\pm0.05$ \\
		Standard - Energy & $0.05\pm0.05$ \\
		CNN - Energy & $0.05\pm0.05$ \\
		\bottomrule
	\end{tabular}
	\vfill
\end{center}

\end{table}

Output features of each reconstruction method are tested independently as well as in a combined set.
The resulting receiver-operating curves and feature importances are shown in Fig.~\ref{fig:roc_curve} and on the right of Tab.~\ref{tab:importances}.
Apart from the method test case, the RF is not capable of significantly distinguishing simulation from experimental data, boosting confidence in the baseline simulation and the reconstruction methods.
Both reconstruction methods provide reasonable data/MC agreement which indicates a certain level of robustness towards potential mis-modeling in the simulation.

\subsection{Ice Systematics}
\label{sec:systematics_systematics}

The accurate description of the detector medium is a challenging task for simulations in IceCube~\cite{IceModelFit}.
Ice properties affect how photons scatter and are absorbed in the detector medium and are therefore
a major contributor to systematic uncertainties in the simulation.
In order to find astrophysical neutrino sources, the impact of these uncertainties on the angular reconstruction must be well understood.

Parameters studied in the context of this paper are the scattering and absorption coefficients of the ice as well as the parameterization of the hole-ice.
These are the main sources of systematic uncertainty in IceCube.
The hole-ice is the refrozen column of ice in which the DOM strings are embedded. 
After deployment of the detector strings, the melted water in the drill holes refreezes.
The refrozen hole-ice has different optical properties with increased absorption and scattering coefficients
due to enclosed bubbles and dust particles~\cite{HoleIce}.
The accurate description of the hole-ice in the simulation is one of the key challenges.
Four systematic datasets are obtained by varying the values of the baseline simulation.

\begin{figure}
  \begin{subfigure}[b]{\textwidth}
    \includegraphics[width=\textwidth, keepaspectratio]{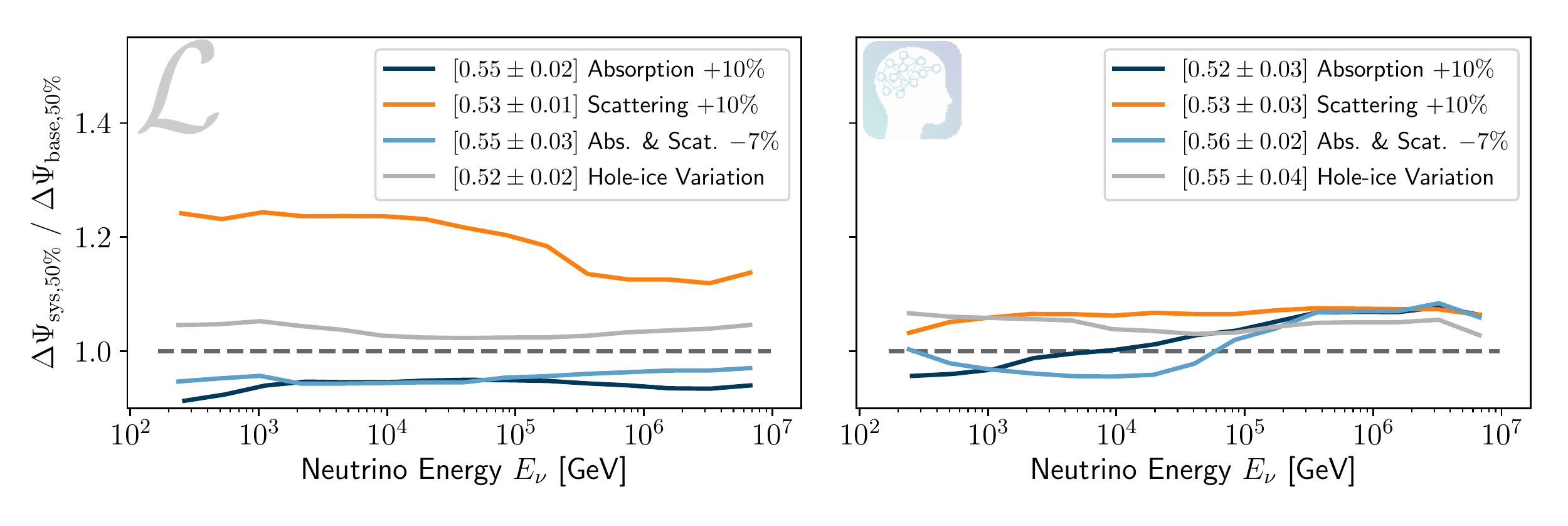}
    \caption{Variation of Ice Properties}
    \label{fig:ice_sys}
  \end{subfigure}
  
  \begin{subfigure}[b]{\textwidth}
    \includegraphics[width=\textwidth, keepaspectratio]{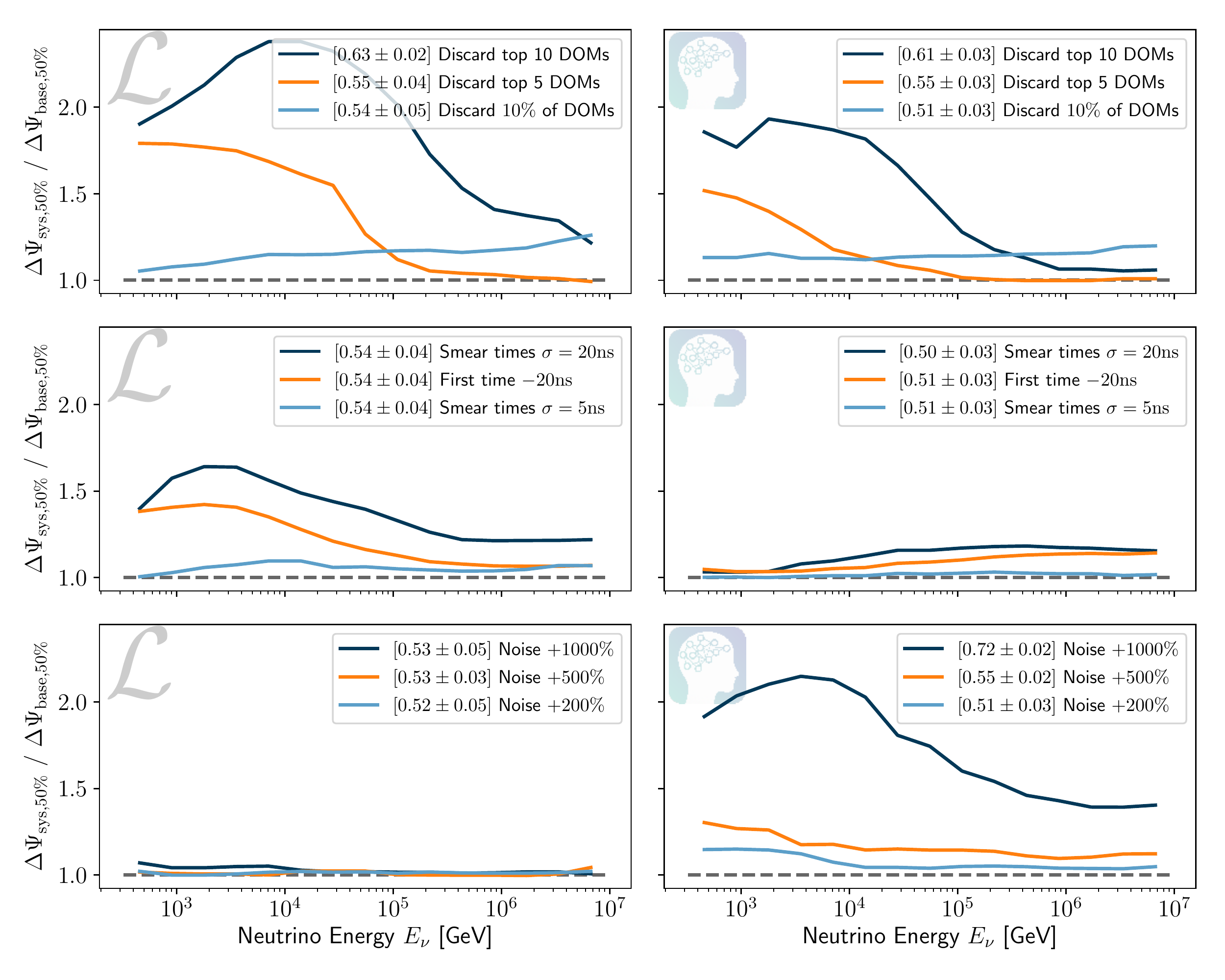}
    \caption{Pulse Modifications}
    \label{fig:pulse_sys}
  \end{subfigure}
  \caption{The median angular resolution is compared between the baseline simulation~$\Delta \Psi_\text{base, 50\%}$ and the systematic variation~$\Delta \Psi_\text{sys, 50\%}$ for the standard reconstruction method (left) and the CNN-based method (right). A ratio close to 1.0 indicates a robust reconstruction method that is insensitive to the varied systematic parameter. For each systematic variation, the data/MC test via the RF (Section~\ref{sec:systematics_agreement}) is performed. The resulting AUC for each variation is shown in brackets in the legend.
}
  \label{fig:robustness_test}
\end{figure}

In Fig.~\ref{fig:ice_sys}, the robustness of the angular resolution towards these systematic variations is investigated.
The ratio of angular resolution for the baseline versus the systematic variation is plotted as a function of neutrino energy. 
A ratio of $1.0$ indicates that the systematic variation has no effect on the angular resolution.
Ratios above 1.0 mean that the systematic variation has a negative effect on the directional reconstruction increasing the opening angle in comparison to the baseline.
In contrast, systematic variations with ratios less than 1.0 improve the angular resolution with respect to the baseline.
The spread in ratios across the different variations is a measure for the robustness of the reconstruction method.
Since the exact values of these parameters are unknown, an ideal reconstruction method should be insensitive to these variations, i.e. produce the same results and therefore have a ratio close to $1.0$. 


Increasing the scattering coefficients will result in more diffuse photons arriving at the photo-multipliers and hence complicate the reconstruction.
As a result, the angular resolution decreases for both reconstruction methods in comparison to the baseline simulation as shown in Fig.~\ref{fig:ice_sys}.
In contrast, reducing the scattering coefficients or increasing the absorption results in an improved angular resolution for the standard reconstruction method.
Overall, the spread across different ice systematics is larger for the standard method than for the CNN-based reconstruction.
With a relative resolution change up to about $5\%$ for the CNN in comparison to $24\%$ for the standard reconstruction method, the CNN seems to be less sensitive to the varied ice-parameters, indicating that the inclusion of the systematic datasets in early stages of the training had a positive effect.

\subsection{Pulse Modifications}
\label{sec:systematics_pulses}

In addition to the varied ice properties in Section~\ref{sec:systematics_systematics}, further unknown sources of systematics may exist.
While the effect of these unknown systematics cannot be directly studied, 
the simulation can be altered in various ways to investigate how a reconstruction method reacts to certain
modifications.
In this study, several modifications are performed to the measured pulses to understand the reconstruction's dependence on single DOMs, its sensitivity to timing uncertainties, and the effect of an increased noise rate.

The reconstruction's dependence on single DOMs is investigated by discarding these prior to the reconstruction.
Entries in the input data tensor~$X$, which correspond to discarded DOMs, are filled with zeros.
Three different variations are tested: discarding the top 5 and top 10 DOMs with the most collected charge and randomly discarding 10\% of all DOMs.
The top panel of Fig.~\ref{fig:pulse_sys} shows the results of this test.
The more light a DOM captures, the more information it carries for the reconstruction task.
Thus, high-charge DOMs (DOMs with the most collected charge in an event) are crucial to the reconstruction.
However, if the reconstruction relies solely, or to a big extent, on a single high-charge DOM, it is more susceptible to mis-modelling in the simulation.
It is challenging to accurately describe local ice properties around a DOM, whereas averages over many DOMs are much more robust.
As shown in the top panel in Fig.~\ref{fig:pulse_sys}, both reconstruction methods have a similar energy-dependent behaviour.
Discarding the highest-charge DOMs negatively impacts the angular resolution at lower energies, while it becomes less important at higher energies.
This is due to the fact that saturated DOMs and overly bright DOMs are automatically excluded from the reconstruction.
This applies to both reconstruction methods which are therefore, by construction, fairly robust to this systematic at higher energies.
For lower energetic events, the number of hit and saturated DOMs reduces and hence the exclusion of the $n$ highest-charge DOMs becomes more and more relevant
as the fraction of removed hit DOMs increases.
Randomly discarding DOMs (light blue curve in top panel) results in a slightly worse angular resolution over the complete energy range with a minimal energy-dependence.
While both reconstruction methods are similarly affected by the exclusion of DOMs, the standard method is more sensitive to single high-charge DOMs, which is driven by the underlying Poisson likelihood.

Unknown systematic uncertainties in the simulation might also affect the arrival times of photons at the photomultipliers.
To investigate this, the pulse times are smeared by a Gaussian with a width of $\SI{5}{ns}$ and $\SI{20}{ns}$.
In comparison, IceCube's timing resolution is on the order of a few ns~\cite{IceCubePMT, DetectorPaper}. 
Moreover, the time of the first pulse at a DOM is particularly important for directional reconstruction, since it is likely to come from a photon that has scattered the least, i.e. carries a lot of information on its origin.
By shifting it by $\SI{20}{ns}$, it can be tested how much the reconstruction relies on this time.
The results are plotted in the middle panel of Fig.~\ref{fig:pulse_sys}.
Smearing the pulse times by $\SI{5}{ns}$, which is larger than IceCube's resolution, barely has an effect on the angular resolution.
Interestingly, the standard method seems to rely more on timing information at lower energies than the CNN-based method.
At energies around a few TeV and lower, the variations to the pulse times have almost no effect on the CNN angular reconstruction.
This indicates that the CNN-based method is not using the timing information to the full extent possible.
Timing information is more important for the reconstruction of low-energetic cascades as opposed to high-energetic cascades for which the charge asymmetry is a good measure of the cascade direction.
This might explain why the standard reconstruction method achieves significantly better resolution at lower energies.

Neural networks can be manipulated to produce wrong results by adding subtle noise to images through evolutionary algorithms or gradient ascent~\cite{FoolingNeuralNetworksOriginal, FoolingNeuralNetworks}.
This vulnerability is of great concern in security crucial applications such as autonomous driving.
While explicit attacks are not a concern for the application in IceCube, potential mis-modeling in the noise simulation could impose a vulnerability. 
Instead of explicitly attempting to manipulate the CNN, the effect of an increased overall noise rate in the detector is studied.
For this study, the noise rate is increased by $\SI{200}{\percent}$, $\SI{500}{\percent}$, and $\SI{1000}{\percent}$.
Note that the values studied here are purely hypothetical and much larger than actual uncertainties on the noise rate, which are on the order of only a few percent.
The noise rate in IceCube is well monitored and stable~\cite{DetectorPaper}.
As previously mentioned, the standard method internally relies on a Poisson likelihood which is driven by high-charge DOMs.
A uniformly increased noise rate therefore barely has any effect.
The standard reconstruction method is robust towards the increased noise rate.
In contrast, the CNN is sensitive to this modification.
In agreement with the previous findings, it seems that the CNN-based method is less focused on single DOMs, but more so on the overall distribution.
Increasing the noise rate to the hypothetical case of an additional $+1000\%$ therefore has an increased impact worsening the angular resolution by a factor up to 2.3.
Realistic variations of the noise rate on the order of a few percent do not have an impact on the angular resolution.

The variations tested here cannot by any means cover all potential systematics in the simulation.
Neural networks can interpolate well on the training data, but should be used with caution when extrapolating.
It is hard to predict how potential mis-modeling in the simulation will affect the reconstruction method.
While this might seem problematic, there are multiple ways to ensure the correctness of deep learning-based approaches.
The neural network can be trained to be robust towards certain variations by including these in the training process, similar to the systematic datasets for the ice properties.
The difficulty lies in compiling an exhaustive list of reasonable variations.
Apart from this, the data/MC test from Section~\ref{sec:systematics_agreement} is an additional safety check.
The data/MC test is performed for every systematic variation of the simulation data and the resulting AUC scores are shown in brackets in the legends of Fig.~\ref{fig:robustness_test}.
Whenever the systematic variation gets large enough to significantly affect the angular resolution, the AUC also increases.
Increasing the noise rate by $\SI{1000}{\percent}$ results in an AUC of $0.72 \pm 0.02$.
Hence, the RF classifier is clearly able to distinguish between data and simulated events.
This increases the confidence that unaccounted systematics will be detected by the data/MC test, if they have significant impact on the reconstruction values.



\section{Discussion on Architecture Limitations}
\label{sec:limitations}


As described in Section~\ref{sec:data}, the input data dimensionality must be reduced and 
transformed in order to be utilized by a standard CNN architecture.
Hence, the measured pulses are summarized in nine input features per DOM.
This effectively removes the time dimension and results in a loss of information.
Quantifying the impact of this information loss as well as the relative importance of each of the
input features require extensive tests which are not further pursued in the context of this paper.
Results shown in Section~\ref{sec:systematics_pulses} do, however, indicate that there are deficiencies.
The neural network is less sensitive than the standard reconstruction method to time-based variables
as shown in the middle panel of Fig.~\ref{fig:pulse_sys}.
This could mean that the neural network is not able to exploit the necessary timing information with the provided data input format and network architecture.

%
%

Further limitations are possible due to the choice of network architecture.
A standard CNN architecture is chosen while only adding minor modifications as detailed in Section~\ref{sec:architecture}.
More advanced CNN architectures 
exist that could be applied.
For this work, no extensive hyper-parameter searches are performed. 
A more suitable and optimized CNN architecture will therefore likely result in further performance gains. 
Apart from CNN architectures, other types of neural networks may be investigated.
Recurrent neural networks are a natural choice given the data representation as a pulse series in time.
Graph neural networks as utilized in Ref.~\cite{NERSC_GNN} can handle the geometry of the detector more adequately. 
Nonetheless, none of the above mentioned architectures are capable of exploiting all available domain knowledge (knowledge that is specific to the application at hand).
The underlying physics of the neutrino interaction are invariant under translation and rotation in space.
The CNN architecture attempts to utilize the translational symmetry through the application of convolutions.
However, due to the detector layout and to the inhomogeneities of the dust concentration in the ice, 
these symmetries are only approximately valid in the measured data.
Translational invariance in time is utilized to a certain degree by the use of relative timing as opposed to absolute timing information.
Available domain knowledge extends beyond the previously mentioned symmetries, 
but it cannot be directly incorporated into the network architecture.
Instead, the neural network relies on learning these relations in a data driven approach which limits its capabilities.
In contrast, methods based on maximum likelihood estimation can include domain knowledge in the likelihood prescription.

\section{Conclusion}
\label{sec:conclusion}

The CNN-based reconstruction method presented in this paper is a versatile tool that complements and in certain parameter spaces improves on state-of-the-art reconstruction methods in IceCube.
It is robust and insensitive to reasonable systematic variations.
A data-MC agreement test via a multivariate classifier can be employed to detect potential unknown and unaccounted systematics, further increasing confidence in the developed method.
The presented method is therefore a viable alternative to traditional reconstruction methods in IceCube.
Due to the significantly improved pointing resolution for cascade-like events above $\SI{10}{TeV}$, 
its application can considerably improve IceCube's sensitivity in the southern sky~\cite{CascadePaperMike7yrs}.
In addition, the CNN's speed and robustness, the low number of outliers and mis-reconstructed events, as well as the available per-event uncertainty estimate make this method attractive for the application in the real-time system.

Despite its success, the CNN-based reconstruction has its limitations.
In the current implementation, the IceCube detector has to be divided into three sub-arrays, 
which are processed independently and only combined at a later stage.
The employed convolutional layers do not optimally use available timing information and
cannot naturally handle the geometry of the detector,
which limits the application of convolutions along the z-axis for the DeepCore sub-arrays.
In addition, the convolution in $x$~and $y$~directions over the main array are only approximately valid due to the distortions in the detector grid.
Similarly, the convolution in $z$-direction is an approximation.
Although the underlying physics of a neutrino interaction are invariant under a translation in space,
the measured light yield in the detector will break this symmetry due to the detector response and inhomogeneity of the dust concentration in the ice.
Apart from translational invariance, other symmetries and prior information exist, which are yet to be exploited.

Future developments will have to overcome these limitations.
The field of geometric deep learning~\cite{GeometricDeepLearning} can help in the correct description of IceCube's irregular geometry.
However, in order to become competitive to the most advanced maximum-likelihood estimation (MLE) methods,
the exploitation of invariances and prior information must be extended. 
Tailored reconstruction methods that combine the benefits of deep learning and MLE must be developed.





\appendix

\section{Neural Network Architecture and Training Details}
\label{sec:appendix}

Details of the neural network architecture and training procedure are provided in this section.
The chosen architecture is inspired by common CNN architectures such as AlexNet~\cite{ImageNetBreakthrough}, ResNet~\cite{ResidualNets}, VGG~\cite{VGG}, and Inception~\cite{GoogleNetInception}.
In addition, the number of layers, filters, and kernel shapes are chosen to strike a balance between computational complexity and expressiveness of the model.
Compared to the previously mentioned CNN architectures
which have between 6 and 138 million free parameters,
the CNN presented here with roughly 6 million parameters is small for the given data complexity.
In combination with the large training dataset, this results in a reduced necessity for regularization as described in Section~\ref{sec:architecture_regularization}.

\begin{table}
  \caption{Details on the number and shape of convolution kernels (filters) and downsampling via max-pooling are provided for the convolutional layers over the main IceCube array (left), upper DeepCore (top right), and lower DeepCore (lower right). Convolutions over the main IceCube array use aligned hexagonally shaped kernels as described in Section~\ref{sec:architecture_hex}. The provided filter and pooling shapes correspond to the dimensions of the orthogonal grid. All layers apply residual additions and use a dropout rate of $d_1$.}
  \label{tab:architecture}
    \begin{center}
        \renewcommand*{\arraystretch}{1.0}
        \begin{tabular}{llll}
            \toprule
            Layer & \# Filters & Filter Shape & Pooling \\
            \midrule
            Hex. 1     & 10  & [5, 5, 7]  & --         \\
            Hex. 2     & 10  & [5, 5, 7]  & --         \\
            Hex. 3     & 20  & [5, 5, 7]  & [1, 1, 2]  \\
            Hex. 4     & 20  & [5, 5, 7]  & --         \\
            Hex. 5     & 20  & [3, 3, 17] & --         \\
            Hex. 6     & 20  & [5, 5, 7]  & --         \\
            Hex. 7     & 20  & [5, 5, 7]  & --         \\
            Hex. 8     & 20  & [3, 3, 17] & --         \\
            Hex. 9     & 80  & [5, 5, 7]  & [2, 2, 2]  \\
            Hex. 10    & 80  & [3, 3, 5]  & --         \\
            Hex. 11    & 80  & [3, 3, 5]  & --         \\
            Hex. 12    & 80  & [3, 3, 5]  & --         \\
            Hex. 13    & 80  & [3, 3, 5]  & --         \\
            Hex. 14    & 80  & [3, 3, 5]  & --         \\
            Hex. 15    & 100 & [3, 3, 5]  & [2, 2, 2]  \\
            Hex. 16    & 100 & [3, 3, 5]  & --         \\
            Hex. 17    & 100 & [3, 3, 5]  & --         \\
            Hex. 18    & 100 & [3, 3, 5]  & [2, 2, 2]  \\
            Hex. 19    & 100 & [1, 1, 3]  & --         \\
            Hex. 20    & 100 & [1, 1, 1]  & --         \\
            \bottomrule
            \vspace{19pt}
        \end{tabular}
        \hfill
        \renewcommand*{\arraystretch}{1.0}
        \begin{tabular}{llll}
            \toprule
            Layer & \# Filters & Filter Shape & Pooling \\
            \midrule
            upper 1  & 80  & [1, 9] & --     \\
            upper 2  & 80  & [1, 9] & --     \\
            upper 3  & 100 & [1, 9] & [1, 2] \\
            upper 4  & 100 & [1, 7] & --     \\
            upper 5  & 100 & [1, 7] & --     \\
            upper 6  & 100 & [1, 7] & [1, 2] \\
            upper 7  & 100 & [1, 5] & --     \\
            upper 8  & 100 & [1, 5] & --     \\
            \midrule
            lower 1     & 80  & [1, 9] & --     \\
            lower 2     & 80  & [1, 9] & --     \\
            lower 3     & 100 & [1, 9] & [1, 2] \\
            lower 4     & 100 & [1, 9] & --     \\
            lower 5     & 100 & [1, 9] & --     \\
            lower 6     & 100 & [1, 9] & [1, 2] \\
            lower 7     & 100 & [1, 9] & --     \\
            lower 8     & 100 & [1, 9] & --     \\
            lower 9     & 100 & [1, 9] & [1, 2] \\
            lower 10    & 100 & [1, 9] & --     \\
            lower 11    & 100 & [1, 9] & --     \\
            lower 12    & 100 & [1, 9] & [1, 2] \\
            lower 13    & 100 & [1, 1] & --     \\
            lower 14    & 100 & [1, 1] & --     \\
            \bottomrule
        \end{tabular}
        \vfill
    \end{center}
\end{table}

\begin{table}
  \caption{The number of nodes per layer, usage of residual additions (Res. Add.), and the applied activation function for the fully-connected layers of the prediction (left) and uncertainty (right) sub-networks are shown. A dropout rate of $d_3$ is applied to all but the output layer. The precursor architecture as has been applied in Ref.~\cite{CascadePaperMike7yrs} does not use layer 2.}
  \label{tab:architecture_output}

  \begin{center}
    \begin{tabular}{lllll}
    \toprule
        Layer & \# Nodes & Res. Add. & Activation \\ \midrule
        1     & 300        & False & elu \\
        2     & 300        & True  & elu \\
        3     & 6          & True  & --    \\
      \bottomrule
    \end{tabular}
  \hfill
    \begin{tabular}{lllll}
    \toprule
        Layer & \# Nodes & Res. Add. & Activation \\ \midrule
        1     & 300        & False & elu \\
        2     & 300        & True  & elu \\
        3     & 6          & True  & abs    \\
      \bottomrule
    \end{tabular}
  \end{center}
\end{table}

\begin{table}
  \caption{Eight training steps as detailed below are performed to train the CNN. A batch size of 32 is used throughout the entire training procedure. The loss target specifies with respect to which sub-networks (prediction/uncertainty or both) the loss function~$\text{L}_{\text{scalar}}$ is optimized. See Section~\ref{sec:architecture_architecture} for the definition of the dropout rates~$d_i$.}
  \label{tab:training_steps}

  \begin{center}
    \begin{tabular}{lllllllll}
      \toprule
      \textbf{Training Step}           & 1 & 2 & 3 & 4 & 5 & 6 & 7 & 8\\ 
      \textbf{Num. Iterations}         & 225500  & 554500  &  309500 & 335500  & 14000  & 6000  & 14000 & 119500 \\
      \textbf{Loss Function}           & MSE & MSE & MSE & MSE & GL & GL & GL & GL       \\
      \textbf{Loss Target}                  & both & both & both & both & unc & unc & both & both\\
      \textbf{Learning Rate}           & 1e-3 & 5e-4 & 1e-4 & 5e-5 & 1e-5 & 1e-5 & 1e-5 & 1e-5 \\
      \textbf{Dropout} $\mathbf{d_0}$  & 0.05 & 0.05 & 0.05 & 0.05 & 0.01 & 0.01 & 0.01 & 0.01 \\
      \textbf{Dropout} $\mathbf{d_1}$  & 0.00 & 0.00 & 0.00 & 0.00 & 0.00 & 0.00 & 0.00 & 0.00 \\
      \textbf{Dropout} $\mathbf{d_2}$  & 0.30 & 0.10 & 0.00 & 0.00 & 0.00 & 0.00 & 0.00 & 0.00 \\
      \textbf{Dropout} $\mathbf{d_3}$  & 0.20 & 0.00 & 0.00 & 0.00 & 0.00 & 0.00 & 0.00 & 0.00 \\
      \textbf{Weight Azimuth}          & 1.0  & 1.0  & 1.0  & 1.0  & 1.0  & 1.0  & 1.0  & 1.0 \\
      \textbf{Weight Zenith}           & 1.0  & 1.0  & 1.0  & 1.0  & 1.0  & 1.0  & 1.0  & 1.0 \\
      \textbf{Weight Energy}           & 1.0  & 1.0  & 1.0  & 1.0  & 1.0  & 10.0  & 10.0  & 5.0 \\
      \textbf{Weight Dir. X}           & 3.0  & 3.0  & 3.0  & 3.0  & 3.0  & 3.0  & 3.0  & 3.0 \\
      \textbf{Weight Dir. Y}           & 3.0  & 3.0  & 3.0  & 3.0  & 3.0  & 3.0  & 3.0  & 3.0 \\
      \textbf{Weight Dir. Z}           & 3.0  & 3.0  & 3.0  & 3.0  & 3.0  & 3.0  & 3.0  & 3.0 \\
      \textbf{Include Sys.}            & True & True & True & False & False & False & False & False \\
      \bottomrule
    \end{tabular}
  \end{center}
\end{table}

\FloatBarrier

\acknowledgments


The IceCube collaboration acknowledges the significant contributions to this manuscript from Mirco H{\"u}nnefeld.
The  authors  gratefully acknowledge the support from the following agencies  and  institutions:
USA {\textendash} U.S. National Science Foundation-Office of Polar Programs,
U.S. National Science Foundation-Physics Division,
U.S. National Science Foundation-EPSCoR,
Wisconsin Alumni Research Foundation,
Center for High Throughput Computing (CHTC) at the University of Wisconsin{\textendash}Madison,
Open Science Grid (OSG),
Extreme Science and Engineering Discovery Environment (XSEDE),
Frontera computing project at the Texas Advanced Computing Center,
U.S. Department of Energy-National Energy Research Scientific Computing Center,
Particle astrophysics research computing center at the University of Maryland,
Institute for Cyber-Enabled Research at Michigan State University,
and Astroparticle physics computational facility at Marquette University;
Belgium {\textendash} Funds for Scientific Research (FRS-FNRS and FWO),
FWO Odysseus and Big Science programmes,
and Belgian Federal Science Policy Office (Belspo);
Germany {\textendash} Bundesministerium f{\"u}r Bildung und Forschung (BMBF),
Deutsche Forschungsgemeinschaft (DFG),
Helmholtz Alliance for Astroparticle Physics (HAP),
Initiative and Networking Fund of the Helmholtz Association,
Deutsches Elektronen Synchrotron (DESY),
and High Performance Computing cluster of the RWTH Aachen;
Sweden {\textendash} Swedish Research Council,
Swedish Polar Research Secretariat,
Swedish National Infrastructure for Computing (SNIC),
and Knut and Alice Wallenberg Foundation;
Australia {\textendash} Australian Research Council;
Canada {\textendash} Natural Sciences and Engineering Research Council of Canada,
Calcul Qu{\'e}bec, Compute Ontario, Canada Foundation for Innovation, WestGrid, and Compute Canada;
Denmark {\textendash} Villum Fonden and Carlsberg Foundation;
New Zealand {\textendash} Marsden Fund;
Japan {\textendash} Japan Society for Promotion of Science (JSPS)
and Institute for Global Prominent Research (IGPR) of Chiba University;
Korea {\textendash} National Research Foundation of Korea (NRF);
Switzerland {\textendash} Swiss National Science Foundation (SNSF);
United Kingdom {\textendash} Department of Physics, University of Oxford.



%
%
%
%
%
%
%
%
\bibliographystyle{JHEP_mod}
\bibliography{references}

\end{document}